\documentclass[useAMS,usenatbib]{mn2e}
\usepackage{color}
\usepackage{multirow}
\usepackage{graphicx}
\usepackage{amssymb} 
\def\aap{A\&A}
\def\aapr{A\&AR}
\def\apj{ApJ}
\def\apjs{ApJS}
\def\apjl{ApJL}

\def\mnras{MNRAS}
\def\aj{AJ}

\def\prd{Phys. Rev. D}

\begin{document}

\title[Variation of fundamental constants: contraints from BOSS-CMASS $\xi(s)$]
{
The clustering of galaxies in the SDSS-III Baryon Oscillation
Spectroscopic Survey: 
constraints on the time variation of fundamental
constants from the large-scale two-point correlation function
}
\author[C.G. Sc\'occola et al.]
{\parbox[t]{\textwidth}{
Claudia G.~Sc\'occola$^{1,2}$\thanks{E-mail: scoccola@iac.es},
Ariel G. S\'anchez$^{3}$,
J.~A.~Rubi\~no-Mart{\'\i}n$^{1,2}$,
R.~G\'enova-Santos$^{1,2}$,
R.~Rebolo$^{1,2}$,
A.~J.~Ross$^{4}$,
W.~J.~Percival$^{4}$,
M.~Manera$^{4}$,
D.~Bizyaev$^{5}$,
J.~R.~Brownstein$^{6}$,
G.~Ebelke$^{5}$,
E.~Malanushenko$^{5}$,
V.~Malanushenko$^{5}$,
D.~Oravetz$^{5}$,
K.~Pan$^{5}$,
D.~P.~Schneider$^{7,8}$,
A.~Simmons$^{5}$
}
\vspace*{6pt} \\ 
$^{1}$ Instituto de Astrof{\'\i}sica de Canarias (IAC), C/V{\'\i}a L\'actea, s/n, La Laguna, Tenerife, Spain. \\
$^{2}$ Dpto. Astrof{\'\i}sica, Universidad de La Laguna (ULL), E-38206 La Laguna, Tenerife, Spain.\\
$^{3}$ Max-Planck-Institut f\"ur extraterrestrische Physik, Postfach 1312, Giessenbachstr., 85748 Garching, Germany.\\ 
$^{4}$ Institute of Cosmology \& Gravitation, University of Portsmouth, Dennis Sciama Building, Portsmouth PO1 3FX, UK.\\
$^{5}$ Apache Point Observatory, P.O. Box 59, Sunspot, NM 88349-0059, USA.\\ 
$^{6}$ Department of Physics and Astronomy, University of Utah, Salt Lake City, UT 84112, USA.\\
$^{7}$ Department of Astronomy and Astrophysics, Pennsylvania State University, University Park, PA 16802, USA.\\
$^{8}$ Institute for Gravitation and the Cosmos, The Pennsylvania State University, University
Park, PA 16802, USA.\\
}

\date{Submitted to MNRAS}
\maketitle
\begin{abstract}
We obtain constraints on the variation of the fundamental constants
from the full shape of the redshift-space correlation function of a
sample of luminous galaxies drawn from the Data Release 9 of the
Baryonic Oscillations Spectroscopic Survey. We combine this
information with additional data from recent cosmic microwave
background, baryon acoustic oscillations and $H_0$ measurements. We
focus on possible variations of the fine structure constant $\alpha$
and the electron mass $m_e$ in the early universe, and study the
degeneracies between these constants and other cosmological
parameters, such as the dark energy equation of state parameter
$w_{\rm DE}$, the massive neutrinos fraction $f_\nu$, the effective
number of relativistic species $N_{\rm eff}$, and the primordial
helium abundance $Y_{\rm He}$.  In the case when only one of the
fundamental constants is varied, our final bounds are $\alpha /
\alpha_0 = 0.9957_{-0.0042}^{+0.0041}$ and $m_e /(m_e)_0 =
1.006_{-0.013}^{+0.014}$.  For the joint variation of both fundamental
constants, our results are $\alpha / \alpha_0 =
0.9901_{-0.0054}^{+0.0055}$ and $m_e /(m_e)_0 = 1.028 \pm 0.019$.  The
variations of $\alpha$ and $m_e$ from their present values affects the
bounds on other cosmological parameters.  Although when $m_e$ is
allowed to vary our constraints on $w_{\rm DE}$ are consistent with a
cosmological constant, when $\alpha$ is treated as a free parameter we
find $w_{\rm DE} = -1.20 \pm 0.13$; more than 1 $\sigma$ away from its
standard value. When $f_\nu$ and $\alpha$ are allowed to vary
simultaneously, we find $f_\nu < 0.043$ (95\% CL), implying a limit of
$\sum m_\nu < 0.46 \, {\rm eV}$ (95\% CL), while for $m_e$ variation,
we obtain $f_\nu < 0.086$ (95\% CL), which implies $\sum m_\nu < 1.1
\, {\rm eV}$ (95\% CL).  When $N_{\rm eff}$ or $Y_{\rm He}$ are
considered as free parameters, their simultaneous variation with
$\alpha$ provides constraints close to their standard values (when the
$H_0$ prior is not included in the analysis), while when $m_e$ is
allowed to vary, their preferred values are significantly higher.  In
all cases, our results are consistent with no variations of $\alpha$
or $m_e$ at the 1 or 2 $\sigma$ level.
\end{abstract}


\section{Introduction}
\label{sec:intro}

The overwhelming amount of cosmological observations obtained over the
past few years has allowed not only the precise determination of the
parameters of the standard cosmological model but also has provided
plenty of scope to test non-standard physics and cosmological
assumptions such as the constancy of fundamental constants over
cosmological timescales.

The variation of fundamental constants is a prediction of theories
attempting to unify the four interactions in nature, such as string
derived field theories, related brane-world theories and Kaluza-Klein
theories \citep[see][and references
  therein]{Uzan2003,Garcia-Berro2007}.  Substantial work have been
devoted to constrain such variations using cosmological observations
\citep{Rahmani2012,Coc2012,Levshakov2012,Menegoni2012,Landau_Scoccola2010}.
Unifying theories predict the variation of all coupling constants,
being all variations related in general to the rolling of a scalar
field. In this paper we adopt a phenomenological approach and analyse
the possible variation of the fine structure constant $\alpha$ and of
the electron mass $m_e$ between the recombination epoch and the
present time, without assuming any theoretical model.

The cosmic microwave background (CMB) is a powerful tool to study the
early universe. The acoustic oscillations present in the CMB power
spectrum are also imprinted through the baryons on the large-scale
structure (LSS) power spectrum \citep{Eisenstein1998,Meiksin1999}. The
correlation function $\xi(s)$ is the Fourier transform of the latter,
and the oscillation structure appears there as a single peak whose
position is related to the sound horizon at the drag redshift
\citep{Matsubara2004}.  The ongoing Baryonic Oscillation Spectroscopic
Survey \citep[BOSS,][]{Dawson2012} is a part of Sloan Digital Sky
Survey-III \citep[SDSS-III][]{Eisenstein2011} and is aimed at
obtaining redshifts for $1.5 \times 10^6$ massive galaxies out to
$z=0.7$ over an area of 10,000 deg$^2$. BOSS is designed to measure
the baryon acoustic oscillations (BAO) signal to probe the expansion
history of the universe. This information places complementary
constraints on the variation of fundamental constants.
A high redshift galaxy sample from BOSS Data Release 9 (DR9),
  denoted CMASS, is constructed through a set of colour-magnitude cuts
  designed to select a roughly volume-limited sample of massive,
  luminous galaxies \citep[][Padmanabhan et al. in
    prep.]{Eisenstein2011}.
The clustering properties of the BOSS CMASS sample
have been analysed in detail in a recent series of papers
\citep{Anderson2012,Manera2012,Reid2012,Ross2012,Samushia2012,Sanchez2012,Tojeiro2012}.

The position of the peak in the correlation function of galaxies can
place constraints on the variation of fundamental constants. Moreover,
the full shape of the correlation function provides additional
information that can break degeneracies, since some parameters vary
the full shape, while others affect only the position and height of
the BAO peak.  We use the full shape of the correlation function of
BOSS-CMASS galaxies presented in \citet{Sanchez2012}, in combination
with CMB observations, to place constraints on the time variation of
fundamental constants in the early universe. We focus on possible
variations in the fine structure constant, $\alpha$, and the electron
mass, $m_e$, at the recombination epoch.  Strictly speaking, the
acoustic fluctuations in the baryons are frozen in at the drag epoch
rather than at last scattering \citep{Hu_Sugiyama1996}. This dynamical
decoupling of the baryons from the photons occurs nevertheless near
recombination. Therefore, we assume that the values of the fundamental
constants are the same throughout this epoch, though they can differ
from their current values.  We analyze the degeneracies with the basic
cosmological parameters, as well as with others, such as the dark
energy equation of state, the neutrino mass, the effective number of
relativistic species, and the primordial helium abundance.

Limits on the present rate of variation of $\alpha$ and $\mu =
m_e/m_p$ (where $m_p$ is the proton mass) are provided by atomic
clocks
\citep{Prestage1995,Sortais2001,Bize2003,Marion2003,Fischer2004,Peik2004}. Data
from the Oklo natural fission reactor \citep{Damour1996,Fujii2000} and
half-lives of long lived $\beta$ decayers \citep{Olive2004} allow to
constrain the variation of fundamental constants at $z \simeq
1$. Absorption systems in the spectra of high-redshift quasars put
additional constraints at different redshifts. The method is based on
the measurement of the separation between spectral lines in doublets
and multiplets, whose dependence on the constants vary among different
species \citep[see for
  example][]{Webb1999,Webb2001,Murphy2003,Agafonova2011,Kanekar2012,Wendt2012}. Although
the limits imposed by CMB and LSS are less stringent than the previous
ones, they are important because they refer to earlier times.

The paper is organized as follows. In Section~\ref{sec:methodology} we
show how the correlation function depends on the values of $\alpha$
and $m_e$ at the recombination epoch.  We describe the datasets used
to place constraints on the variation of the fundamental constants and
the statistical method performed. The modelling of the correlation
function is also summarized. Section~\ref{sec:results} presents the
results for different parameter spaces. Conclusions are outlined in
Section~\ref{sec:conclusions}.

\section{Methodology} 
\label{sec:methodology}

We performed a statistical analysis to constrain the variation of
$\alpha$ and $m_e$ at the recombination epoch, together with other
cosmological parameters varied. In this Section we describe the
datasets used to obtain our results. Then, we summarize the
modelling of the correlation function, and present a brief
explanation of why the correlation function is an effective observable to
constrain the variation of fundamental constants. Finally, we describe the 
statistical analysis employed.

\subsection{Data}

In this paper, we use the full shape of the large-scale two-point
correlation function $\xi(s)$ of the BOSS-CMASS galaxy sample,
computed in \citet{Sanchez2012}.  This function was computed using the
first spectroscopic data release of BOSS \citep[Data Release 9,
  DR9,][]{Ahn2012}.  The galaxy target selection of BOSS is divided in
two separate samples, named LOWZ and CMASS, covering different
redshifts \citep{Eisenstein2011,Padmanabhan_inprep,Dawson2012}.  This
selection is based on photometric observations done with the dedicated
2.5-m Sloan Telescope \citep{Gunn2006}, located at Apache Point
Observatory in New Mexico, using a drift-scanning mosaic CCD camera
\citep{Gunn1998} that produces $ugriz$ images \citep{Fukugita1996}.
The CMASS sample is constructed on the basis of $gri$ colour cuts
designed to select luminous galaxies such that they constitute an
approximately complete galaxy sample down to a limiting stellar mass
\citep{Maraston_inprep}.
Spectra of the LOWZ and CMASS samples are obtained using the
double-armed BOSS spectrographs, which are significantly upgraded from
those used by SDSS-I/II \citep{York2000}, covering the wavelength
range 3600~\AA{} to 10000~\AA{} with a resolving power of 1500 to 2600
\citep{Smee2012}.
Spectroscopic redshifts are measured using the minimum-$\chi^2$
template-fitting procedure described in \citet{Aihara2011}, with
templates and methods updated for BOSS data as described in
\citet{Bolton_inprep}.  \citet{Anderson2012} present a detailed
description of the construction of the catalogue for LSS studies based
on this sample, and the calculation of the completeness of each sector
of the survey mask, i.e., the areas of the sky covered by a unique set
of spectroscopic tiles.

To constrain high-dimensional parameter spaces, it is necessary to
combine the CMASS $\xi(s)$ with other datasets. In our analysis, we
use the WMAP 7-year temperature and temperature-polarization power
spectra \citep{Larson2011} and the results from the South Pole
Telescope \citep[SPT,][]{Keisler2011}. The latter provide information
on the structure of the acoustic peaks in the CMB power spectrum up to
multipoles $\ell \simeq 3000$.  As discussed in \citet{Keisler2011},
in the multipole range covered by SPT ($650 \leq \ell \leq 3000$), the
CMB power spectrum contains a non-negligible contribution from
secondary anisotropies, while for $\ell \lesssim 650$ the main
contribution arises from primary anisotropies. We follow the approach
of \citet{Keisler2011} to account for the secondary anisotropies, by
including the contribution from the Sunyaev-Zel'dovich (SZ) effect and
the emission from foreground galaxies (including both a clustered and
a Poisson source contribution), using templates whose amplitudes are
considered as nuisance parameters and marginalized over. The WMAP-SPT
combination is referred to as our ``CMB'' dataset.

We also use information from other clustering measurements in the form
of constraints on the position of the baryon acoustic peak from
independent analyses. We use the results of \citet{Beutler2011} which
are based on measurements of the large-scale correlation function of
the 6dF Galaxy Survey \citep[6DFGS,][]{Jones2009} and the 2\% distance
measurement obtained by \citet{Padmanabhan2012} and \citet{Xu2012}
from the application of an updated version of the reconstruction
technique proposed by \citet{Eisenstein2007} to the clustering of
galaxies from the final SDSS-II LRG sample
\citep{York2000,Eisenstein2011}. The results of these analyses are
combined in the ``BAO'' dataset.

Lastly, we consider a Gaussian prior on the Hubble parameter based on
the latest Hubble Space Telescope (HST) observations of H$_0 = 73.8
\pm 2.4$ km\,s$^{-1}$\,Mpc$^{-1}$ \citep{Riess2011}.

The aforementioned datasets are used in different combinations to
check the consistency of the obtained bounds. Firstly, we use the CMB
data alone, and then combine it with the CMASS correlation
function. In the end, we combine the four datasets to obtain our final
constraints.

We do not consider supernovae (SNs) type Ia data because the light
curves of the SNs are obtained assuming that the fundamental constants
have their present values at the observing redshift. However, since we
are investigating a possible time evolution in the value of $\alpha$
and $m_e$, and the SNs are at considerably high redshift
\citep[$0.7<z<1.4$ for the high-$z$ sample of][]{Conley2011}, we
cannot neglect the possibility that the constants have a different
value at those times. In fact, several studies aiming at measuring the
value of $\alpha$ at high redshift using quasar absorption systems do
not conclusively exclude the variation of fundamental constants at
those redshifts
\citep{Webb1999,Webb2011,Murphy2003,Murphy2004,King2012}. Therefore,
to be conservative, we choose not to consider the supernovae datasets
in our analysis.

\subsection{Model for the correlation function}

We follow \citet{Sanchez2012} and model the shape of the large-scale
correlation function, $\xi(s)$, by applying the following
parametrization:
\begin{equation}
 \xi(s) = b^2 \left[\xi_{\rm L}(s)\otimes {\rm e}^{-(k_{\star}s)^2} 
+ A_{\rm MC} \,\xi'_{\rm L}(s)\,\xi^{(1)}_{\rm L}(s) \right], 
\label{eq:xi_model}
\end{equation}
where the symbol $\otimes$ denotes a convolution, and the bias factor
$b$, mode-coupling amplitude $A_{\rm MC}$, and the smoothing length
$k_{\star}$ are considered as free parameters and marginalized over.
Here $\xi'_{\rm L}$ is the derivative of the linear correlation
function $\xi_{\rm L}$, and $\xi^{(1)}_{\rm L}(s)$ is defined by
\begin{equation}
 \xi_{\rm L}^{(1)}(s) \equiv \hat{s} \cdot \nabla^{-1}\xi_{\rm L}(s)
=\frac{1}{2\pi^2}\int P_{\rm L}(k)\,j_1(ks)k\,{\rm d}k ,
\label{eq:xi1}
\end{equation}
with $j_{\rm 1}(y)$ denoting the spherical Bessel function of order 1.

The parametrization of equation~\ref{eq:xi_model} was first proposed
by \citet{Crocce2008} and is based on renormalized perturbation theory
\citep[RPT,][]{Crocce2006}. \citet{Sanchez2008} compared this model
against the results of an ensemble of large volume N-body simulations
\citep[L-BASICC-II,][]{Angulo2008}, and showed that it provides an
accurate description of the full shape of the correlation function,
including also the effects of bias and redshift-space
distortions. This parametrization has been applied to obtain
constraints on cosmological parameters from clustering measurements
from various galaxy samples \citep{Sanchez2009,Beutler2011,Blake2011}.

As in ~\citet{Sanchez2012}, we restrict the comparison of the model of
equation~\ref{eq:xi_model} and the BOSS-CMASS correlation function to
$40 < s < 200 \, h^{-1}\, {\rm Mpc}$, and assume a Gaussian likelihood
function.

\subsection{Effects on the full-shape of $\xi(s)$}

During the recombination epoch, the ionization fraction is determined
by the balance between photoionization and recombination. The most
important effects of changes in $\alpha$ and $m_e$ during this epoch
are due to their influence upon Thomson scattering cross section
$\sigma_T = 8\pi \hbar^2 \alpha^2 / 3 m_e^2 c^2$ and the binding
energy of hydrogen $B_1 = \frac{1}{2} \alpha^2 m_e c^2$.  The
ionization history is more sensitive to $\alpha$ than to $m_e$ because
of the $B_1$ dependence on these constants. The main result is the
shift of the epoch of recombination to higher $z$ as $\alpha$ or $m_e$
increases, which corresponds to a smaller sound horizon.  These
effects are imprinted into the matter power spectrum through the
transfer function. Hence, they also affect the galaxy correlation
function. Consequently, if at recombination $\alpha$ or $m_e$ had a
value higher than the present one, the peak in the correlation
function would appear at smaller scales, since the position of the
peak is related to the size of the sound horizon at the drag epoch.

\begin{figure}
\centerline{\includegraphics[angle=-90,width=0.6\textwidth]{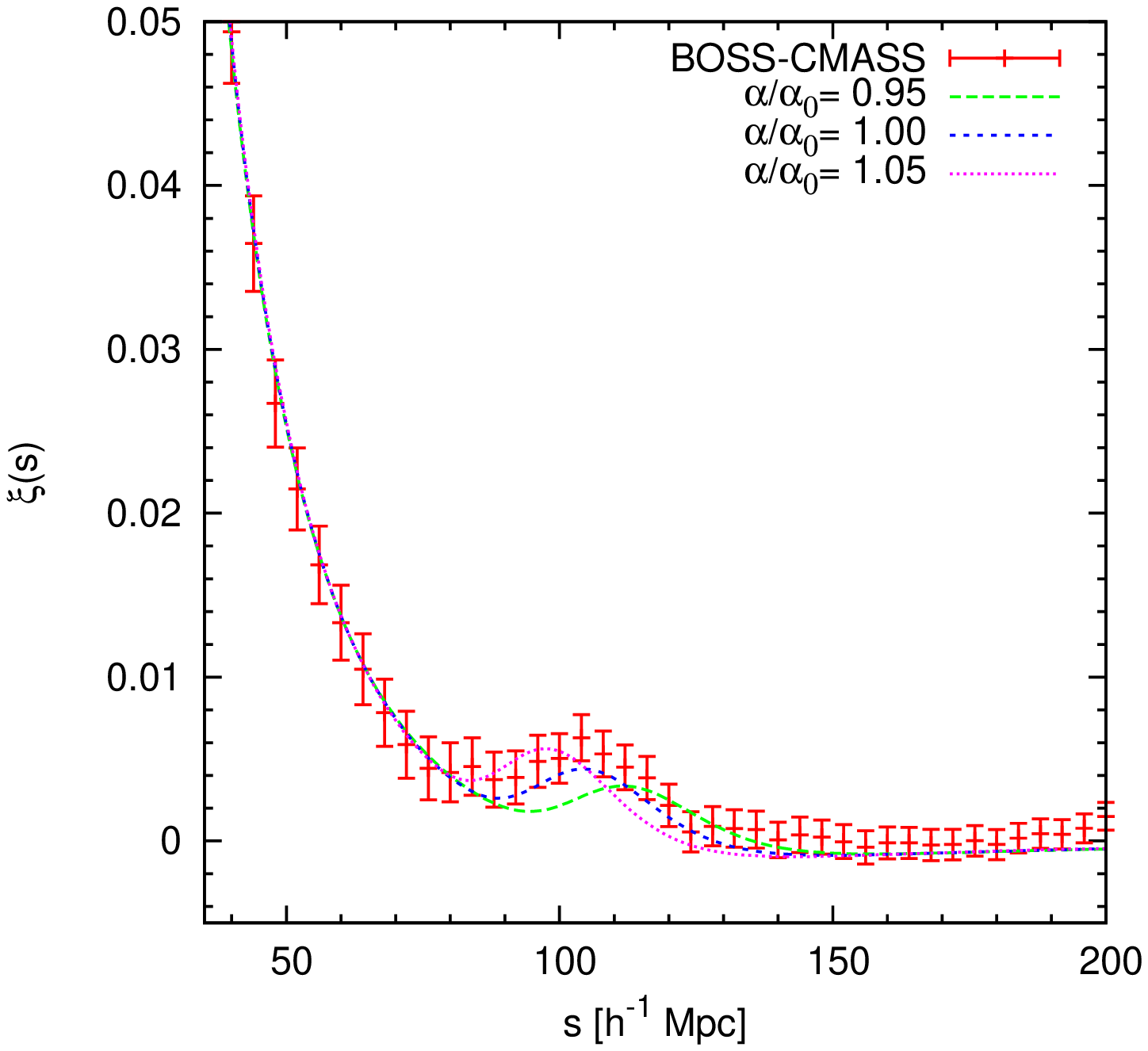}}
\centerline{\includegraphics[angle=-90,width=0.6\textwidth]{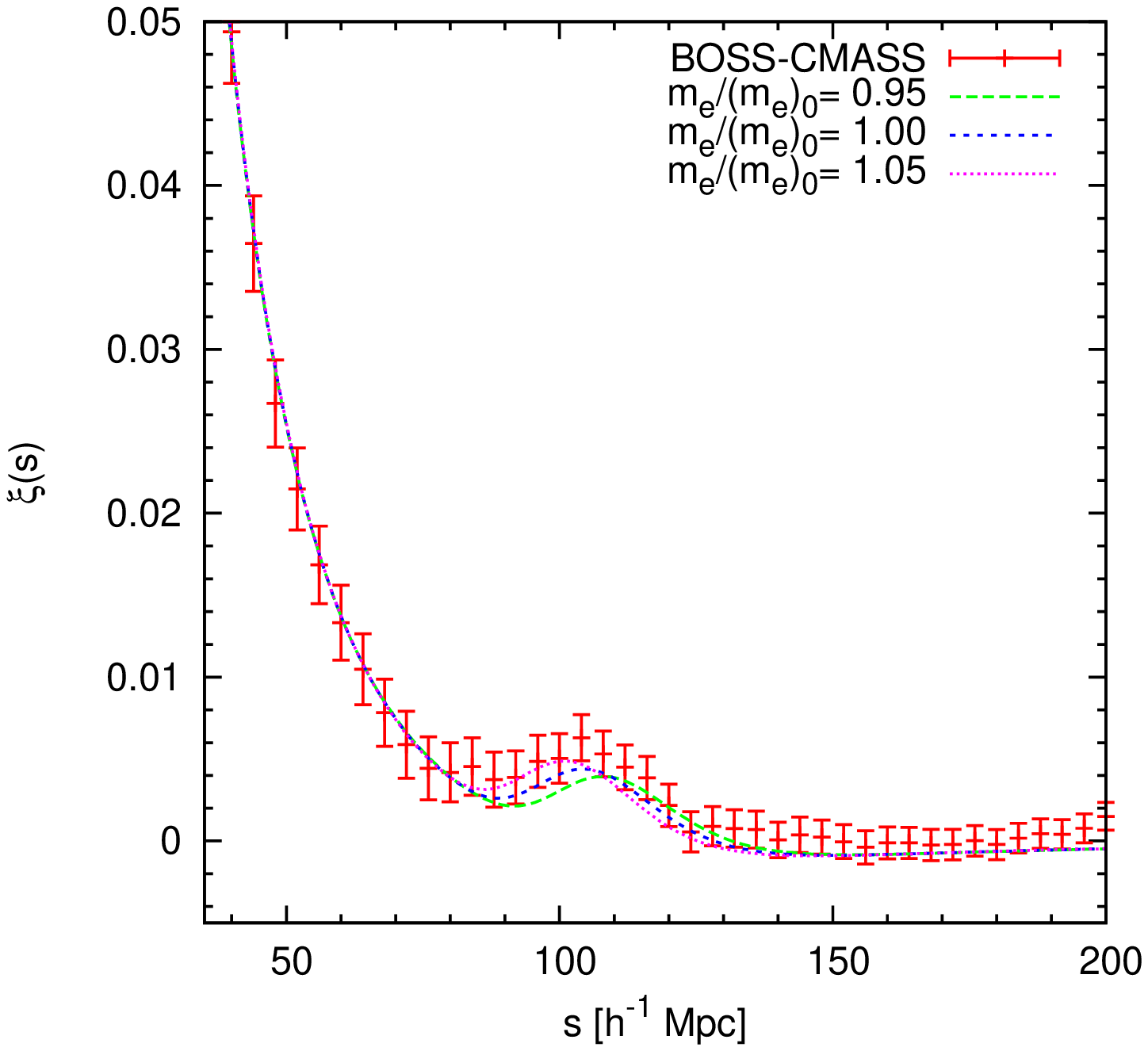}}
\caption{Effects on the two-point correlation function of a 5\%
  variation in $\alpha$ (upper panel), and in $m_e$ (lower panel) at
  recombination time, with respect to their present values. Other
  parameters are kept fixed}. Data points are the measurement of the
BOSS-CMASS two-point correlation function from \citet{Sanchez2012}.
\label{fig:CF_alfa_emasa}
\end{figure}

Fig.~\ref{fig:CF_alfa_emasa} shows the effect of a variation in
$\alpha$ and in $m_e$, during the recombination epoch, that we should
expect in the correlation function of CMASS galaxies, in a flat
universe with cosmological parameters $(\omega_{\rm b}, \, \omega_{\rm
  dm}, \, \tau, \, h, \, n_{\rm s}) = (0.0221, \, 0.1145, \, 0.696, \,
0.962)$. We show the prediction for different values of the
fundamental constants: their present value, and a variation of $\pm
5\%$ with respect to their values today. Data points are the
spherically averaged redshift-space two-point correlation function of
the full CMASS sample presented in \citet{Sanchez2012}.

The changes in the correlation function due to variations in $\alpha$
are larger than the changes in $m_e$ for a given relative variation of
their value. Both constants affect the position and height of the
peak, leaving the rest of the curve unchanged. This effect can break
degeneracies with other cosmological parameters that affect the full
shape of the correlation function, such as the dark energy equation of
state. Other parameters also affect the position of the
  peak. This leads to degeneracies between the fundamental constants
  and other cosmological parameters, such as $\Omega_{\rm m}$, as we
  will see in Sections~\ref{Sec:alpha} and~\ref{Sec:me}. The combination of CMASS data with
  different datasets helps to break those degeneracies.

\subsection{Statistical Analysis}

We perform our statistical analysis by exploring the parameter spaces
with Monte Carlo Markov chains generated with the {\sc CosmoMC} code
\citep{Lewis-Bridle2002}, which uses the Boltzmann code {\sc camb}
\citep{Lewis2000} and {\sc recfast} \citep{Seager1999} to compute the
CMB power spectra.  In order to be able to study more general models
in which the dark energy component is different from the cosmological
constant, we use a generalized version of {\sc camb} which supports
values of the dark energy equation of state beyond the phantom divide,
$w_{\rm DE} < -1$ \citep{Fang2008}.  We modified these codes to
include the variation in $\alpha$ and $m_e$ at recombination as
described in \citet{Landau2008}. Additional modifications from
\citet{Keisler2011} are included to compute the likelihood of the SPT
dataset.  The dependence on the fundamental constants of the detailed
physics relevant in the recombination process is described in
\citet{Scoccola2008}. Nevertheless, we emphasize that such description
is done in terms of a modification to the effective 3-level atom model
which is used in {\sc Recfast} \citep{Wong2008}. Additional physical
processes \citep[see e.g.,][for a review]{Rubino2008,Fendt2009} are
effectively treated using a correction function inside {\sc Recfast
  v1.5} \citep{Rubino2010}. This function has, to first order, a
negligible dependence on the standard cosmological parameters, but
also on non-standard parameters such as $Y_{\rm He}$ or $N_{\rm eff}$
\citep{ShawChluba2011}. Although it has not been demonstrated
explicitly, this correction function is also expected to have a small
dependence on $\alpha$ and $m_e$. Thus, we will use in this paper the
correction function as it appears in {\sc Recfast v1.5}. However, for
future CMB experiments with higher sensitivities in the damping tail
of the angular power spectrum, a more detailed and complete treatment
of the recombination problem might be relevant \citep{CosmoRec}.

We consider a spatially-flat cosmological model with adiabatic density
fluctuations, described by the following parameters, which define the
$\Lambda$CDM model:
\begin{equation}
P = \left( \omega_{\rm b}, \omega_{\rm dm}, \Theta, \tau, A_{\rm s}, n_{\rm s} \right)
\label{eq:standard_parameters}
\end{equation}
where $\omega_{\rm b} = \Omega_{\rm b} h^2$ is the baryon density,
$\omega_{\rm dm} = \Omega_{\rm dm} h^2$ is the dark matter density,
both in units of the critical density; $\Theta$ gives the ratio of the
comoving sound horizon at decoupling to the angular diameter distance
to the surface of last scattering; $\tau$ is the reionization optical
depth; and $A_{\rm s}$ and $n_{\rm s}$ are the amplitude and spectral
index of the primordial power spectrum of the scalar fluctuations,
respectively, at the pivot wavenumber of $k = 0.05$\,Mpc$^{-1}$.

To this set of cosmological parameters, we add the value of $\alpha$
and/or $m_e$ at the recombination epoch as additional variables. We
introduce them as $\alpha/\alpha_0$ and $m_e/(m_e)_0$, i.e., relative
to their present value, denoted with a $0$ subscript.

Moreover, to constrain possible deviations from the $\Lambda$CDM
model, we study also the cases in which we allow for variations of the
dark energy equation of state $w_{\rm DE}$ and the dark matter
fraction in the form of massive neutrinos, $f_\nu = \Omega_\nu /
\Omega_{\rm dm}$, focusing on the degeneracies with the fundamental
constants studied here. In addition, we analyze the case in which the
effective number of relativistic species is different from its
standard value of $N_{\rm eff} = 3.046$, and the case in which the
primordial helium fraction can have a value different than the
standard one of $Y_{\rm He} = 0.24$.  Finally, we present the
constraints on the parameters of the $\Lambda$CDM+$N_{\rm eff}$ and
$\Lambda$CDM+$Y_{\rm He}$ models in the case of no variation of the
fundamental constants.

We also present constraints on other quantities, derived from the
above parameter set. These are the sum of the neutrino masses, given
by
\begin{equation}
\sum m_\nu = 94.4 \, \omega_{\rm dm} \, f_\nu \, \,  {\rm eV},
\end{equation} 
the dark energy density $\Omega_{\rm DE}$, the total matter density
$\Omega_{\rm m}$, the rms linear perturbation theory variance in
spheres of radius 8$\, h^{-1}$Mpc $\sigma_{\rm 8}$, the age of the
universe $t_{\rm 0}$, the redshift of reionization $z_{\rm re}$, and
the hubble factor $h$.

The parameters $b$, $A_{\rm MC}$, and $k_{\star}$ of the model for the
full shape of the correlation function, and the amplitudes
$D_{3000}^{\rm SZ}$, $D_{3000}^{\rm PS}$, $D_{3000}^{\rm CL}$,
necessary to account for the secondary anisotropies in the SPT power
spectrum, are treated as nuisance parameters, and marginalized over
when presenting our results.

 For most of the cases, we only use the information from the CMASS
 correlation function in combination with our CMB data. However, in
 Sections~\ref{Sec:alpha} and \ref{Sec:me}, we also ran chains using
 the CMASS information alone, to study the degeneracies between
 $\alpha$ and $m_e$ and other cosmological parameters obtained from
 this dataset. In these cases we impose Gaussian priors on
 $\omega_{\rm b}$ and $n_{\rm s}$, obtained from CMB-only results (see
 first column of Table~\ref{tab:al} and Table~\ref{tab:me}).

\section{Results}
\label{sec:results}

We present the constraints obtained for the fundamental constants and
cosmological parameters in each of the cases studied. Errorbars will
indicate 68\% confidence level (CL) unless otherwise stated. In
Section~\ref{Sec:alpha} we study the variation of the fine structure
constant, and do the statistical analysis varying also the
cosmological parameters of the $\Lambda$CDM model. In
Section~\ref{Sec:me} we study the variation of the electron mass
together with the cosmological parameters. Section~\ref{Sec:alpha_me}
investigates the joint variation of $\alpha$ and $m_e$. In
Section~\ref{Sec:alpha_wde} we analyze the constraints in the case
where one of the constants ($\alpha$ or $m_e$) and the dark energy
equation of state $w_{\rm DE}$ can take values that differ from the
standard ones. In Section~\ref{Sec:alpha_fnu} we study the case where
we vary one of the fundamental constants and the massive neutrinos
fraction $f_\nu$. Section~\ref{Sec:alpha_Neff} focusses on the
constraints obtained when the fundamental constants are varied
together with the effective number of relativistic species, $N_{\rm
  eff}$, while Section~\ref{Sec:alpha_YHe} presents the constraints
when the fundamental constants are varied together with the primordial
helium fraction $Y_{\rm He}$. In Appendix~\ref{sec:table_LCDM}, we
give the constraints on the $\Lambda$CDM model for our four datasets,
to facilitate the comparison to the results presented in this paper.

\subsection{Variation of $\alpha$}
\label{Sec:alpha}

In this Section, we extend the $\Lambda$CDM model to include possible
variations in the fine structure constant during recombination. We
present our constraints on $\alpha/\alpha_0$ and the cosmological
parameters.

\begin{figure}
\centerline{\includegraphics[angle=-90,width=0.6\textwidth]{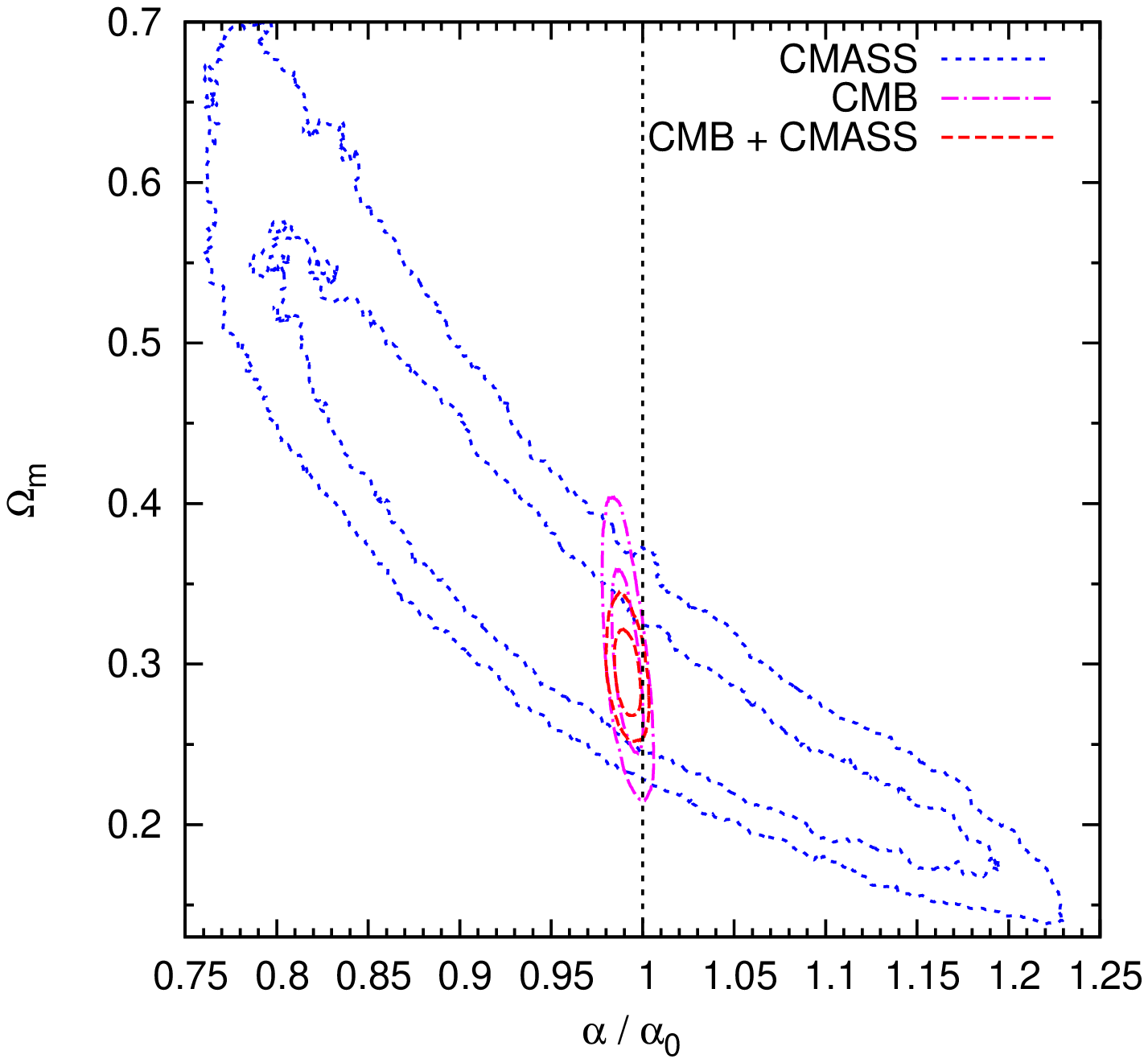}}
\centerline{\includegraphics[angle=-90,width=0.6\textwidth]{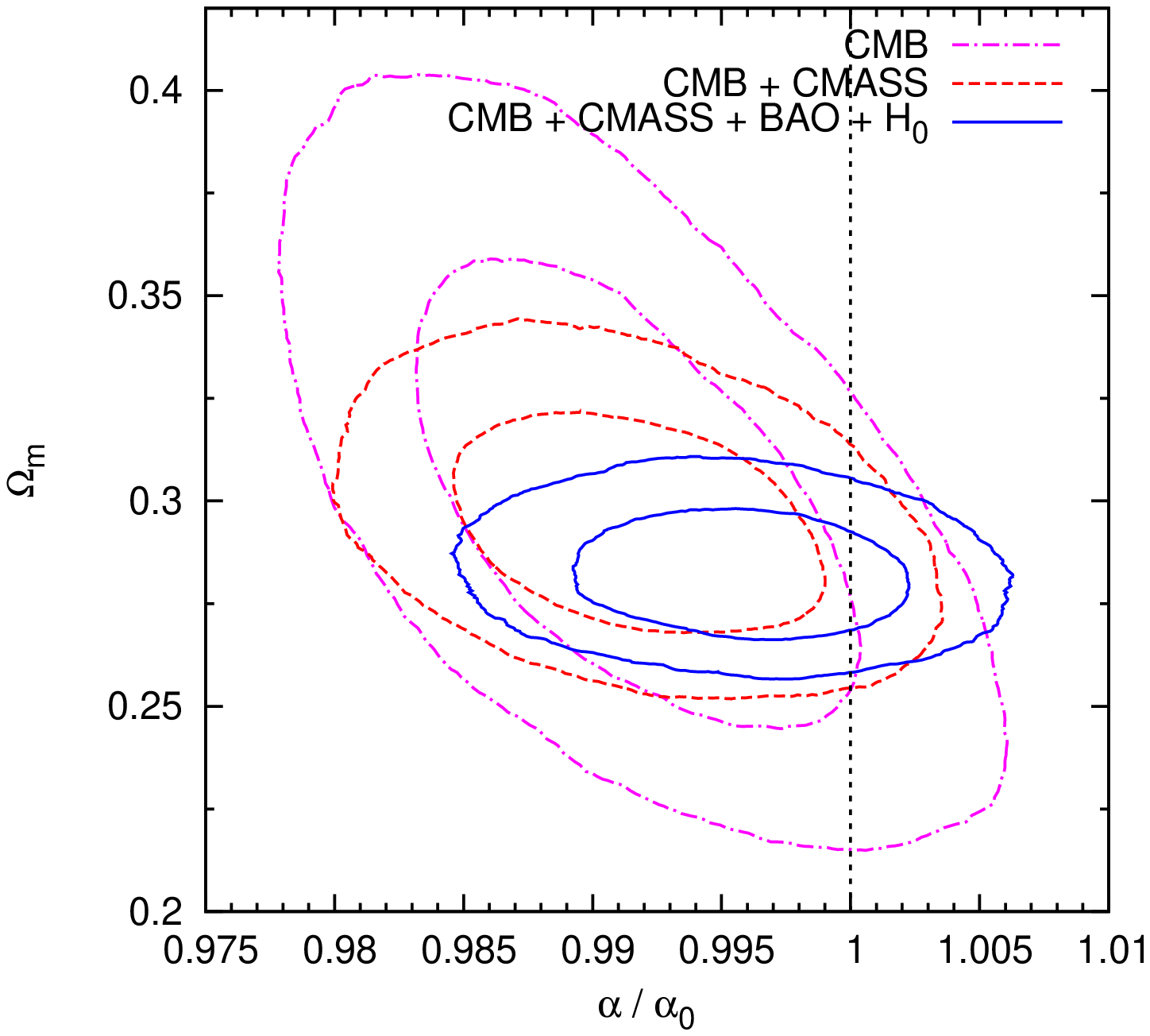}}
\caption{ The marginalized posterior distribution in the $\alpha$ -
  $\Omega_{\rm m}$ plane for the $\Lambda$CDM parameter set extended
  to include the variation of $\alpha$.  The dot-dashed lines show the
  68\% and 95\% contours obtained using CMB information alone.
    The dotted contours, in the upper panel, show the results from CMASS
    correlation function.  The dashed lines correspond to the results
  obtained from the combination of CMB data plus the shape of the
  CMASS $\xi(s)$. The solid lines indicate the results obtained from
  the full dataset combination (CMB+CMASS+BAO+$H_0$).  The
    vertical dotted line corresponds to the $\Lambda$CDM model, with
    $\alpha/\alpha_0=1$}.
\label{fig:2D_al_om_solo-alfa}
\end{figure}

 The upper panel of Fig.~\ref{fig:2D_al_om_solo-alfa} shows the
 two-dimensional marginalized constraints in the $\alpha$ -
 $\Omega_{\rm m}$ plane obtained from the CMB and CMASS datasets in
 isolation.  As specified in Section~\ref{sec:methodology}, when using
 the information of the CMASS correlation function alone we impose
 Gaussian priors on $\omega_{\rm b}$ and $n_{\rm s}$ consistent with
 our CMB-only results.  The constraints obtained from these datasets
 exhibit strong degeneracies.  As these degeneracies constrain
 different combinations of $\alpha$ and $\Omega_{\rm m}$, the
 combination of the two datasets provides tighter constraints on both
 parameters simultaneously.  As can seen in the lower panel of
 Fig.~\ref{fig:2D_al_om_solo-alfa}, the remaining degeneracy between
 these parameters is alleviated when more datasets are added to the
 analysis.  The correlation factors are $-0.62$, $-0.38$, and $-0.13$
 for CMB, CMB+CMASS, and the full dataset combination
 (CMB+CMASS+BAO+$H_0$), respectively.

\begin{figure}
\centerline{\includegraphics[angle=-90,width=0.6\textwidth]{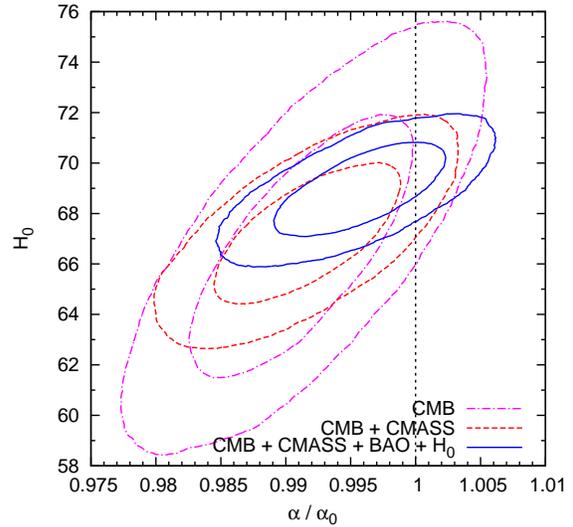}}
\caption{ The marginalized posterior distribution in the $\alpha$ -
  $H_0$ plane for the $\Lambda$CDM parameter set extended to include
  the variation of $\alpha$. The dot-dashed lines show the 68\% and
  95\% contours obtained using CMB information alone. The dashed lines
  correspond to the results obtained from the combination of CMB data
  plus the shape of the CMASS $\xi(s)$. The solid lines indicate the
  results obtained from the full dataset combination
  (CMB+CMASS+BAO+$H_0$).  }
\label{fig:2D_al_H0_solo-alfa}
\end{figure}

In Fig.~\ref{fig:2D_al_H0_solo-alfa} we show the constraints in the
$\alpha$ - $H_0$ plane. $H_0$ is better constrained when additional
datasets are included in the analysis. The value of the HST prior on
$H_0$ lies almost 2 $\sigma$ from the value obtained for $H_0$ from
the CMB+CMASS dataset, being the latter considerably smaller. Hence,
when the HST prior on $H_0$ is taken into account, the obtained value
for the Hubble parameter is increased. Due to the degeneracy with
$\alpha$, the contours are shifted in the parameter space when the
$H_0$ prior is included, and the value of the fine structure constant
is increased, being closer to its present value.

In Table~\ref{tab:al} we present the constraints on the variation of
$\alpha$ and the cosmological parameters. The constraint on $\alpha$
from CMB data alone is $\alpha / \alpha_0 = 0.9914 \pm 0.0055$.  The
CMB-only constraints on $\Omega_{\rm m}$ are considerably degraded by
the inclusion of $\alpha$ as a free parameter, with $\Omega_{\rm
  m}=0.303^{+0.037}_{-0.036}$ (compare to the $\Lambda$CDM value of
$\Omega_{\rm m}$ in Table~\ref{tab:LCDM} of the Appendix).  When we
add the information encoded in the full shape of the CMASS $\xi(s)$,
we find $\alpha / \alpha_0 = 0.9917 \pm 0.0046$ and $\Omega_{\rm
  m}=0.295^{+0.017}_{-0.018}$.  In the case that all datasets are used
in the analysis, the constraints are
$\alpha=0.9957_{-0.0042}^{+0.0041}$ and $\Omega_{\rm m}=0.283\pm
0.010$, which is the same precision as that obtained in the
$\Lambda$CDM model.  The inclusion of the additional datasets produces
only a mild improvement of the bounds on $\alpha$ with respect to
those obtained using CMB information alone.  This result arises
because our CMB dataset covers the high multipole range, with $\ell
\sim 3000$, which imposes strong constraints on this parameter.  The
bounds obtained for the cosmological parameters from the full dataset
are consistent within 1 $\sigma$ with their values in the $\Lambda$CDM
model (see Table~\ref{tab:LCDM}).

\citet{Menegoni2012} present constraints on the variation of $\alpha$
using CMB data including data from SPT and the Atacama Cosmology
Telescope \citep[ACT,][]{Dunkley2011}, both probing the damping regime
of the CMB fluctuations.  By combining this information with the
galaxy power spectrum from the SDSS-DR7 luminous
red galaxy sample \citep{Reid2010} and the HST prior on $H_0$, they
find $\alpha / \alpha_0 = 0.984 \pm 0.005$.  The precision on this
constraint on $\alpha$ is provided mostly by the CMB data, explaining
why our results with CMB data alone have almost the same precision as
theirs. Nevertheless, the CMASS $\xi(s)$ further improves the
precision of the bound. It is important to emphasize, however, that
our results prefer values for $\alpha$ that are closer to its present
value than theirs. Our results are consistent with no
variation of $\alpha$ at 2 $\sigma$ (for the full dataset, they are
just slightly inconsistent at 1 $\sigma$).


\begin{table} 
\centering
  \caption{ The marginalized 68\% allowed regions on the cosmological
    parameters of the $\Lambda$CDM model, adding the variation of the
    fine structure constant, $\alpha$, obtained using different
    combinations of the datasets.  }
    \begin{tabular}{@{}lccc@{}}
    \hline
& \multirow{2}{*}{CMB}  & \multirow{2}{*}{CMB + CMASS} &  CMB + CMASS \\
&                       &                           &   + BAO + $H_0$   \\  
\hline
$\alpha / \alpha_0$  &  $0.9914_{-0.0055}^{+0.0055}$ &  $0.9917_{-0.0046}^{+0.0046}$ &  $0.9957_{-0.0042}^{+0.0041}$ \\[1.5mm]
100$\Theta$  &  $1.0289_{-0.0078}^{+0.0078}$ &  $1.0293_{-0.0065}^{+0.0064}$ &  $1.0353_{-0.0058}^{+0.0057}$ \\[1.5mm]
100$\omega_{\rm b}$  &  $2.207_{-0.043}^{+0.043}$ &  $2.209_{-0.039}^{+0.039}$ &  $2.225_{-0.039}^{+0.038}$ \\[1.5mm]
100$\omega_{\rm dm}$  &  $11.17_{-0.47}^{+0.48}$ &  $11.09_{-0.36}^{+0.36}$ &  $11.19_{-0.34}^{+0.33}$ \\[1.5mm]
$\tau$  &  $0.0880_{-0.0073}^{+0.0064}$ &  $0.0877_{-0.0072}^{+0.0064}$ &  $0.0867_{-0.0072}^{+0.0061}$ \\[1.5mm]
$n_{\rm s}$  &  $0.977_{-0.013}^{+0.013}$ &  $0.977_{-0.013}^{+0.013}$ &  $0.973_{-0.013}^{+0.013}$ \\[1.5mm]
ln$(10^{10}A_{\rm s})$  &  $3.104_{-0.034}^{+0.034}$ &  $3.101_{-0.031}^{+0.032}$ &  $3.095_{-0.030}^{+0.030}$ \\[1.5mm]
$\Omega_{\rm DE}$  &  $0.697_{-0.037}^{+0.036}$ &  $0.705_{-0.017}^{+0.018}$ &  $0.717_{-0.010}^{+0.010}$ \\[1.5mm]
$\Omega_{\rm m}$  &  $0.303_{-0.036}^{+0.037}$ &  $0.295_{-0.018}^{+0.017}$ &  $0.283_{-0.010}^{+0.010}$ \\[1.5mm]
$\sigma_{\rm 8}$  &  $0.815_{-0.023}^{+0.023}$ &  $0.813_{-0.020}^{+0.020}$ &  $0.818_{-0.018}^{+0.018}$ \\[1.5mm]
$t_{\rm 0}$/Gyr  &  $14.12_{-0.26}^{+0.26}$ &  $14.10_{-0.19}^{+0.20}$ &  $13.90_{-0.16}^{+0.17}$ \\[1.5mm]
$z_{\rm re}$  &  $10.8_{-1.2}^{+1.2}$ &  $10.8_{-1.2}^{+1.2}$ &  $10.6_{-1.2}^{+1.2}$ \\[1.5mm]
$h$   &  $0.668_{-0.033}^{+0.033}$ &  $0.672_{-0.018}^{+0.018}$ &  $   0.689_{-0.012}^{+0.012}$ \\[1.5mm]
\hline
\end{tabular}
\label{tab:al}
\end{table}


\subsection{Variation of $m_e$}
\label{Sec:me}

Now we turn to the case in which we extend the $\Lambda$CDM model
  to include possible variations in the electron mass during
  recombination.

\begin{figure}
\centerline{\includegraphics[angle=-90,width=0.6\textwidth]{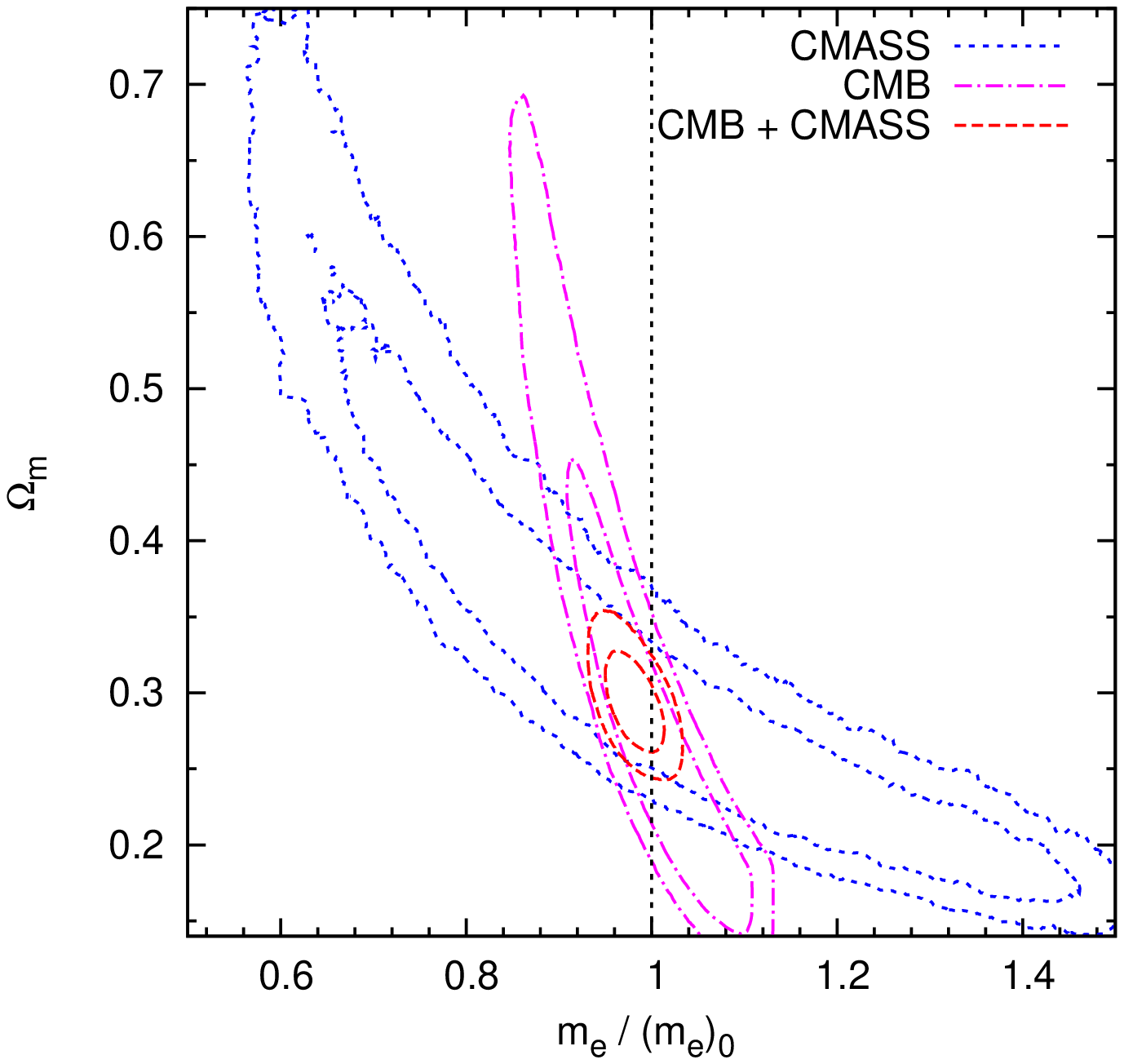}}
\centerline{\includegraphics[angle=-90,width=0.6\textwidth]{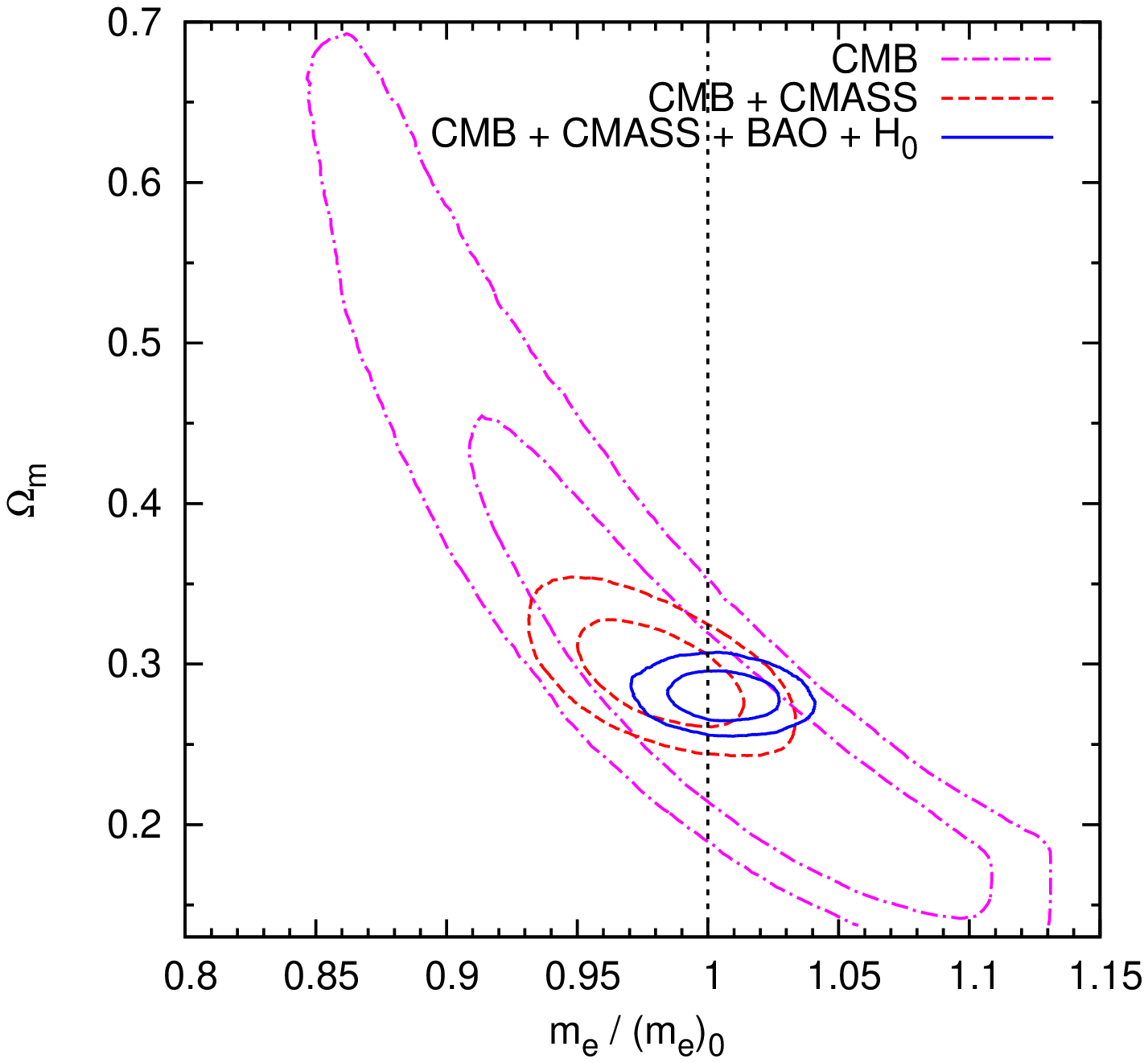}}
\caption{ The marginalized posterior distribution in the $m_e$ -
  $\Omega_{\rm m}$ plane for the $\Lambda$CDM parameter set extended
  to include the variation of $m_e$. The dot-dashed lines show the
  68\% and 95\% contours obtained using CMB information alone.  The
  dotted contours, in the upper panel, show the results from CMASS
  correlation function. } The dashed lines correspond to the results
obtained from the combination of CMB data plus the shape of the CMASS
$\xi(s)$. The solid lines indicate the results obtained from the full
dataset combination (CMB+CMASS+BAO+$H_0$). The vertical dotted line
corresponds to the $\Lambda$CDM model, with $m_e / (m_e)_0=1$.
\label{fig:2D_me_om_solo-emasa}
\end{figure}

The contours in Fig.~\ref{fig:2D_me_om_solo-emasa} show the
two-dimensional marginalized constraints in the $m_e$ - $\Omega_{\rm
  m}$ plane.  The upper panel shows that the constraints obtained from
the CMB and CMASS datasets in isolation exhibit strong degeneracies in
different directions in the parameter space.  The CMB-only constraints
exhibit a strong degeneracy between $m_e$ and $\Omega_{\rm m}$,
causing $\Omega_{\rm m}$ to be poorly constrained by the CMB data
alone, with $\Omega_{\rm m}=0.31^{+0.12}_{-0.11}$.  The inclusion of
the CMASS correlation function improves the constraints on
$\Omega_{\rm m}$ by a factor larger than five to obtain $\Omega_{\rm
  m}=0.295\pm0.021$. This bound is further reduced by a factor of two
when all datasets are included, in which case we find $\Omega_{\rm
  m}=0.280\pm0.010$, having the same precision than in the
$\Lambda$CDM model.

\begin{figure}
\centerline{\includegraphics[angle=-90,width=0.6\textwidth]{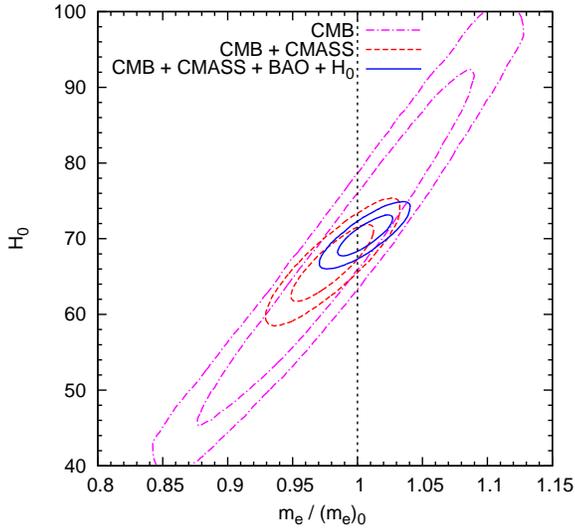}}
\caption{ The marginalized posterior distribution in the $m_e$ -
  $H_0$ plane for the $\Lambda$CDM parameter set extended to
  include the variation of $m_e/(m_e)_0$. The dot-dashed lines show the 68\%
  and 95\% contours obtained using CMB information alone. The
  dashed lines correspond to the results obtained from the combination
  of CMB data plus the shape of the CMASS $\xi(s)$. The solid lines
  indicate the results obtained from the full dataset combination
  (CMB+CMASS+BAO+$H_0$).}
\label{fig:2D_me_H0_solo-emasa}
\end{figure}

In Fig.~\ref{fig:2D_me_H0_solo-emasa} we show the two-dimensional
marginalized constraints in the $m_e$ - $H_0$ plane. When $m_e$ is
allowed to vary, CMB information alone is insufficient to place any
reliable constraint on $H_0$. When the CMASS dataset is added, the
constraint improves noticeably. This dataset breaks the degeneracy
between $H_0$ and $m_e$. When all the datasets are considered, the
value of $H_0$ is increased due to the HST prior, and the value of
$m_e$ is shifted towards its present value.

In Table~\ref{tab:me} we present the constraints on the variation of
$m_e$ and the cosmological parameters. The constraint on $m_e$ from
CMB data alone is $0.989_{-0.069}^{+0.067}$.  When we add the
information encoded in the full shape of the CMASS $\xi(s)$, the bound
is $0.981_{-0.021}^{+0.020}$. In the case that all datasets are used
in the analysis, the constraint is $1.006_{-0.013}^{+0.014}$.  The
precision in the bound is highly improved when we add the information
of the CMASS $\xi(s)$ to the analysis. Our final results are
completely consistent with no variation of $m_e$ within 1 $\sigma$.


\begin{table} 
\centering
  \caption{ The marginalized 68\% allowed regions on the cosmological
    parameters of the $\Lambda$CDM model, adding the variation of the
    electron mass, $m_e$, obtained using different combinations of the
    datasets.  }
    \begin{tabular}{@{}lccc@{}}
    \hline
& \multirow{2}{*}{CMB}  & \multirow{2}{*}{CMB + CMASS} &  CMB + CMASS \\
&                       &                           &   + BAO + $H_0$   \\  
\hline
$m_e /(m_e)_0$  &  $0.989_{-0.069}^{+0.067}$ &  $0.981_{-0.021}^{+0.020}$ &  $1.006_{-0.013}^{+0.014}$ \\[1.5mm]
100$\Theta$  &  $1.033_{-0.051}^{+0.049}$ &  $1.027_{-0.015}^{+0.015}$ &  $1.0448_{-0.0094}^{+0.0097}$ \\[1.5mm]
100$\omega_{\rm b}$  &  $2.20_{-0.17}^{+0.17}$ &  $2.177_{-0.054}^{+0.054}$ &  $2.228_{-0.042}^{+0.042}$ \\[1.5mm]
100$\omega_{\rm dm}$  &  $11.06_{-0.88}^{+0.87}$ &  $10.95_{-0.64}^{+0.63}$ &  $11.67_{-0.54}^{+0.55}$ \\[1.5mm]
$\tau$  &  $0.0850_{-0.0071}^{+0.0063}$ &  $0.0850_{-0.0069}^{+0.0061}$ &  $0.0813_{-0.0066}^{+0.0059}$ \\[1.5mm]
$n_{\rm s}$  &  $0.965_{-0.012}^{+0.012}$ &  $0.9646_{-0.0099}^{+0.0100}$ &  $0.9620_{-0.0098}^{+0.0098}$ \\[1.5mm]
ln$(10^{10}A_{\rm s})$  &  $3.080_{-0.031}^{+0.031}$ &  $3.078_{-0.030}^{+0.031}$ &  $3.089_{-0.029}^{+0.029}$ \\[1.5mm]
$\Omega_{\rm DE}$  &  $0.69_{-0.12}^{+0.11}$ &  $0.705_{-0.021}^{+0.021}$ &  $0.720_{-0.010}^{+0.010}$ \\[1.5mm]
$\Omega_{\rm m}$  &  $0.31_{-0.11}^{+0.12}$ &  $0.295_{-0.021}^{+0.021}$ &  $0.280_{-0.010}^{+0.010}$ \\[1.5mm]
$\sigma_{\rm 8}$  &  $0.799_{-0.074}^{+0.072}$ &  $0.794_{-0.039}^{+0.039}$ &  $0.839_{-0.031}^{+0.032}$ \\[1.5mm]
$t_{\rm 0}$/Gyr  &  $14.1_{-1.6}^{+1.7}$ &  $14.20_{-0.47}^{+0.47}$ &  $13.64_{-0.28}^{+0.28}$ \\[1.5mm]
$z_{\rm re}$  &  $10.3_{-1.4}^{+1.4}$ &  $10.2_{-1.1}^{+1.1}$ &  $10.3_{-1.2}^{+1.2}$ \\[1.5mm]
$h$  &  $0.69_{-0.15}^{+0.15}$ &  $0.668_{-0.032}^{+0.032}$ &  $0.704_{-0.017}^{+0.017}$ \\[1.5mm]
\hline
\end{tabular}
\label{tab:me}
\end{table}


The bounds for the cosmological parameters are consistent within
  1 $\sigma$ with those from the $\Lambda$CDM model. The mean values
  of $\sigma_{\rm 8}$ and $h$ for the CMB+CMASS dataset are somewhat
  smaller than in the $\Lambda$CDM model, but still consistent within
  1 $\sigma$.

\subsection{Joint variation of $\alpha$ and $m_e$}
\label{Sec:alpha_me}

In this Section, we extend the $\Lambda$CDM model to study the joint
variation of $\alpha$ and $m_e$.

\begin{figure}
\centerline{\includegraphics[angle=-90,width=0.6\textwidth]{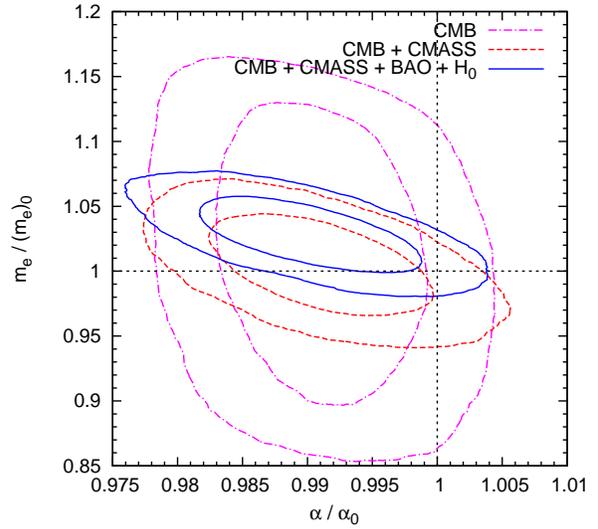}}
\caption{ The marginalized posterior distribution in the $\alpha$ -
  $m_e$ plane for the $\Lambda$CDM parameter set extended to include
  the joint variation of $\alpha$ and $m_e$. The dot-dashed lines show
  the 68\% and 95\% contours obtained using CMB information alone. The
  dashed lines correspond to the results obtained from the combination
  of CMB data plus the shape of the CMASS $\xi(s)$. The solid lines
  indicate the results obtained from the full dataset combination
  (CMB+CMASS+BAO+$H_0$).}
\label{fig:2D_al_me_joint-variation}
\end{figure}

The contours in Fig.~\ref{fig:2D_al_me_joint-variation} show the
two-dimensional marginalized constraints in the $m_e$ - $\alpha$
plane. When the $H_0$ prior is used, the mean value of $\alpha$ is
marginally decreased, while the mean value of $m_e$ is increased. The inclusion
of additional datasets reduces the allowed region in the parameter
space, while increasing the correlation between the fundamental
constants; the correlation factor is $-0.23$, $-0.55$, and $-0.68$ for
CMB, CMB+CMASS, and the full dataset combination
(CMB+CMASS+BAO+$H_0$), respectively.

\begin{figure}
\centerline{\includegraphics[angle=-90,width=0.6\textwidth]{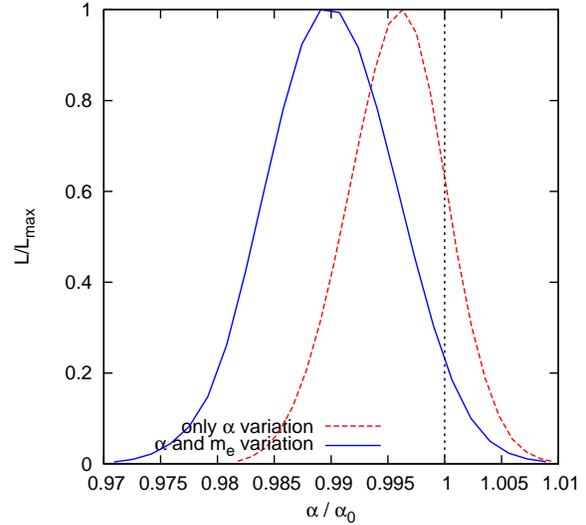}}
\caption{ The marginalized, one-dimensional likelihood distribution
  for the fine structure constant, when only $\alpha$ is allowed to
  vary, and in the case of the joint variation of $\alpha$ and $m_e$.}
\label{fig:1D_alfa}
\end{figure}

 In Table~\ref{tab:al_me} we present the constraints obtained for the
 fundamental constants and the cosmological parameters. When adding
 different datasets, the precision in the determination of $\alpha$
 remains the same, although the mean value is slightly lower when the
 full dataset is used.  For the CMB and CMB+CMASS datasets, the mean
 value of $\alpha$ is almost the same, regardless of $m_e$ being fixed
 to its present value or allowed to vary. For the full dataset, the
 value of $\alpha$ is decreased by 1 $\sigma$ when $m_e$ is also
 allowed to vary (see Fig.~\ref{fig:1D_alfa}).

\begin{table} 
\centering
  \caption{
    The marginalized 68\% allowed regions on the cosmological parameters of the $\Lambda$CDM model, adding the variation of the fine structure constant $\alpha$ and of the electron mass $m_e$,
    obtained using different combinations of the datasets.
}
    \begin{tabular}{@{}lccc@{}}
    \hline
& \multirow{2}{*}{CMB}  & \multirow{2}{*}{CMB + CMASS} &  CMB + CMASS \\
&                       &                           &   + BAO + $H_0$   \\  
\hline
$\alpha / \alpha_0$  &  $0.9909_{-0.0055}^{+0.0055}$ &  $0.9910_{-0.0055}^{+0.0055}$ &  $0.9901_{-0.0054}^{+0.0055}$ \\[1.5mm]
$m_e / (m_e)_0$  &  $1.012_{-0.073}^{+0.073}$ &  $1.004_{-0.025}^{+0.025}$ &  $1.028_{-0.019}^{+0.019}$ \\[1.5mm]
100$\Theta$  &  $1.036_{-0.052}^{+0.051}$ &  $1.031_{-0.015}^{+0.015}$ &  $1.0467_{-0.0094}^{+0.0095}$ \\[1.5mm]
100$\omega_{\rm b}$  &  $2.24_{-0.18}^{+0.18}$ &  $2.216_{-0.059}^{+0.059}$ &  $2.265_{-0.046}^{+0.046}$ \\[1.5mm]
100$\omega_{\rm dm}$  &  $11.29_{-0.92}^{+0.92}$ &  $11.20_{-0.66}^{+0.65}$ &  $11.83_{-0.55}^{+0.56}$ \\[1.5mm]
$\tau$  &  $0.0879_{-0.0074}^{+0.0066}$ &  $0.0877_{-0.0071}^{+0.0067}$ &  $0.0848_{-0.0071}^{+0.0063}$ \\[1.5mm]
$n_{\rm s}$  &  $0.978_{-0.014}^{+0.014}$ &  $0.978_{-0.013}^{+0.013}$ &  $0.977_{-0.012}^{+0.012}$ \\[1.5mm]
ln$(10^{10}A_{\rm s})$  &  $3.105_{-0.035}^{+0.035}$ &  $3.104_{-0.035}^{+0.035}$ &  $3.114_{-0.033}^{+0.033}$ \\[1.5mm]
$\Omega_{\rm DE}$  &  $0.69_{-0.12}^{+0.11}$ &  $0.706_{-0.020}^{+0.020}$ &  $0.719_{-0.010}^{+0.010}$ \\[1.5mm]
$\Omega_{\rm m}$  &  $0.31_{-0.11}^{+0.12}$ &  $0.294_{-0.020}^{+0.020}$ &  $0.281_{-0.010}^{+0.010}$ \\[1.5mm]
$\sigma_{\rm 8}$  &  $0.824_{-0.081}^{+0.078}$ &  $0.820_{-0.043}^{+0.042}$ &  $0.860_{-0.035}^{+0.036}$ \\[1.5mm]
$t_{\rm 0}$/Gyr   &  $14.0_{-1.7}^{+1.7}$ &  $14.04_{-0.47}^{+0.47}$ &  $13.56_{-0.28}^{+0.27}$ \\[1.5mm]
$z_{\rm re}$  &  $10.9_{-1.5}^{+1.5}$ &  $10.8_{-1.3}^{+1.3}$ &  $11.0_{-1.3}^{+1.3}$ \\[1.5mm]
$h$    &  $0.70_{-0.16}^{+0.16}$ &  $0.677_{-0.033}^{+0.033}$ &  $0.708_{-0.017}^{+0.017}$ \\[1.5mm]
\hline
\end{tabular}
\label{tab:al_me}
\end{table}


\begin{figure}
\centerline{\includegraphics[angle=-90,width=0.6\textwidth]{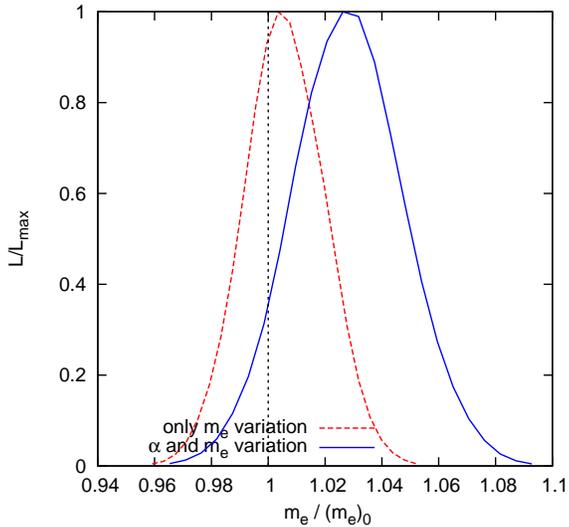}}
\caption{ The marginalized, one-dimensional likelihood distribution
  for the electron mass, when only $m_e$ is allowed to vary, and in
  the case of the joint variation of $\alpha$ and $m_e$.}
\label{fig:1D_emasa}
\end{figure}

The constraints on $m_e$ are significantly improved when adding extra
datasets.  The bound on $m_e$ from CMB-only data is improved by almost
a factor of three when the CMASS dataset is added to the analysis, and
is further improved when all the datasets are considered. The mean
value of $m_e$ is increased when $\alpha$ is also allowed to vary,
compared to the case in which $\alpha$ is fixed to its present
value. For the CMB+CMASS and the full datasets, the increment is more
than 1 $\sigma$ (see Fig.~\ref{fig:1D_emasa} for the full dataset
case).  For the full dataset combination, the mean value of $m_e$
takes its largest value.  Our final bounds, using the full dataset,
are $\alpha / \alpha_0 = 0.9901_{-0.0054}^{+0.0055}$ and $m_e /
(m_e)_0 = 1.028_{-0.019}^{+0.019}$. Both limits are consistent with no
variation of the fundamental constants at the 2 $\sigma$ level.

Most of the cosmological parameters are consistent within 1 $\sigma$
with their $\Lambda$CDM model values for each of the datasets. The
value of $n_{\rm s}$ is larger than in the $\Lambda$CDM model, for all
of the datasets. For the CMB+CMASS and the full datasets, $n_{\rm s}$
is larger by slightly more than 1 $\sigma$. The value of $\sigma_{\rm
  8}$ is slightly larger than its value in the $\Lambda$CDM model, but
consistent within 1 $\sigma$.

\subsection{Variation of fundamental constants and $w_{\rm DE}$ }
\label{Sec:alpha_wde}

Until now, we have assumed that the dark energy component corresponds
to a cosmological constant, with a fixed equation of state specified
by $w_{\rm DE} = -1$. In this Section we explore the constraints on
the value of $w_{\rm DE}$, assumed to be redshift-independent, in the
context of the variation of fundamental constants.

In this study, the dynamical dark energy models are allowed to cross
the so-called phantom divide, $w_{\rm DE} = -1$, to explore models
with $w_{\rm DE} < -1$. In the framework of general relativity, a
single scalar field cannot cross this threshold, since it becomes
gravitationally unstable
\citep{Feng2005,Vikman2005,Hu2005,Xia2008}. Thus, more degrees of
freedom are required, which are difficult to implement in general dark
energy studies. We follow the parametrized post-Friedmann (PPF)
approach of \citet{Fang2008}, as implemented in {\sc camb}, which
provides a simple solution to these problems for models in which the
dark energy component is smooth compared to the dark matter.

\begin{figure}
\centerline{\includegraphics[angle=-90,width=0.6\textwidth]{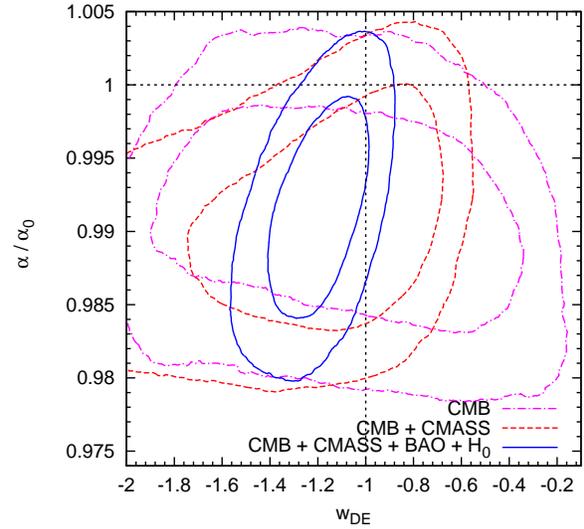}}
\caption{ The marginalized posterior distribution in the $\alpha$ -
  $w_{\rm DE}$ plane for the $\Lambda$CDM parameter set extended to
  include the variation of $\alpha$ and the redshift-independent value
  of $w_{\rm DE}$ as an additional parameter. The dot-dashed lines
  show the 68\% and 95\% contours obtained using CMB information
  alone. The dashed lines correspond to the results obtained from the
  combination of CMB data plus the shape of the CMASS $\xi(s)$. The
  solid lines indicate the results obtained from the full dataset
  combination (CMB+CMASS+BAO+$H_0$). The vertical and horizontal
  dotted lines correspond to the $\Lambda$CDM model, with
  $\alpha/\alpha_0=1$ and $w_{\rm DE}= -1$.}
\label{fig:2D_al_wde_variation_al-wde}
\end{figure}

First, we perform a statistical analysis varying $\alpha$ and $w_{\rm
  DE}$, together with the rest of the cosmological
parameters. Fig.~\ref{fig:2D_al_wde_variation_al-wde} presents the
constraints on the $\alpha$ - $w_{\rm DE}$ plane.  Again, the
inclusion of additional datasets reduces the allowed region in the
parameter space, while increasing the anti-correlation between the
fundamental constants. The correlation factor is $-0.16$, $0.37$, and
$0.54$ for CMB, CMB+CMASS, and the full dataset combination
(CMB+CMASS+BAO+$H_0$), respectively. For the full dataset, the value
$w_{\rm DE} \geq -1$ is excluded at more than 1 $\sigma$.

Table~\ref{tab:al_wde} shows the constraints obtained for $\alpha$,
$w_{\rm DE}$, and the cosmological parameters. The inclusion of more
datasets slightly improves the constraint on $\alpha$ and does
  not appreciably affect the mean value. In the case of $w_{\rm DE}$,
  a large improvement is observed, and the mean value of $w_{\rm DE}$
  shifts towards lower values. Our final bounds, using  the
complete dataset, are $\alpha / \alpha_0 = 0.9915 \pm 0.0048$ and
$w_{\rm DE} = -1.20 \pm 0.13$. Our results are consistent with no
variation of $\alpha$ at the 2 $\sigma$ level, while $w_{\rm DE}$ is
compatible with a cosmological constant also within 2 $\sigma$,
although at 1 $\sigma$, both quantities differ from their
  standard values.


\begin{table} 
\centering
  \caption{ The marginalized 68\% allowed regions on the cosmological
    parameters of the $\Lambda$CDM model, adding the variation of the
    fine structure constant $\alpha$, and the dark energy equation of
    state $w_{\rm DE}$, obtained using different combinations of the
    datasets.  }
    \begin{tabular}{@{}lccc@{}}
    \hline
& \multirow{2}{*}{CMB}  & \multirow{2}{*}{CMB + CMASS} &  CMB + CMASS \\
&                       &                           &   + BAO + $H_0$   \\  
\hline
$w_{\rm DE}$  &  $-1.07_{-0.50}^{+0.48}$ &  $-1.23_{-0.41}^{+0.36}$ &  $-1.20_{-0.13}^{+0.13}$ \\[1.5mm]
$\alpha / \alpha_0$  &  $0.9908_{-0.0054}^{+0.0054}$ &  $0.9907_{-0.0050}^{+0.0049}$ &  $0.9915_{-0.0048}^{+0.0048}$ \\[1.5mm]
100$\Theta$  &  $1.0280_{-0.0078}^{+0.0077}$ &  $1.0277_{-0.0071}^{+0.0070}$ &  $1.0289_{-0.0068}^{+0.0068}$ \\[1.5mm]
100$\omega_{\rm b}$  &  $2.205_{-0.042}^{+0.042}$ &  $2.202_{-0.042}^{+0.042}$ &  $2.200_{-0.040}^{+0.039}$ \\[1.5mm]
100$\omega_{\rm dm}$  &  $11.18_{-0.48}^{+0.48}$ &  $11.22_{-0.47}^{+0.46}$ &  $11.42_{-0.35}^{+0.35}$ \\[1.5mm]
$\tau$  &  $0.0883_{-0.0075}^{+0.0063}$ &  $0.0866_{-0.0073}^{+0.0063}$ &  $0.0857_{-0.0069}^{+0.0064}$ \\[1.5mm]
$n_{\rm s}$  &  $0.978_{-0.014}^{+0.014}$ &  $0.976_{-0.013}^{+0.013}$ &  $0.973_{-0.012}^{+0.013}$ \\[1.5mm]
ln$(10^{10} A_{\rm s})$  &  $3.107_{-0.034}^{+0.034}$ &  $3.104_{-0.032}^{+0.033}$ &  $3.107_{-0.031}^{+0.031}$ \\[1.5mm]
$\Omega_{\rm DE}$  &  $0.69_{-0.14}^{+0.13}$ &  $0.736_{-0.057}^{+0.063}$ &  $0.733_{-0.014}^{+0.014}$ \\[1.5mm]
$\Omega_{\rm m}$  &  $0.31_{-0.13}^{+0.14}$ &  $0.264_{-0.063}^{+0.057}$ &  $0.267_{-0.014}^{+0.014}$ \\[1.5mm]
$\sigma_{\rm 8}$  &  $0.83_{-0.14}^{+0.15}$ &  $0.88_{-0.11}^{+0.12}$ &  $0.882_{-0.045}^{+0.045}$ \\[1.5mm]
$t_{\rm 0}$/Gyr  &  $14.19_{-0.43}^{+0.44}$ &  $14.08_{-0.20}^{+0.20}$ &  $14.04_{-0.18}^{+0.18}$ \\[1.5mm]
$ z_{\rm re}$  &  $10.9_{-1.3}^{+1.3}$ &  $10.7_{-1.2}^{+1.3}$ &  $10.7_{-1.2}^{+1.2}$ \\[1.5mm]
$h$  &  $0.70_{-0.15}^{+0.16}$ &  $0.728_{-0.087}^{+0.098}$ &  $0.715_{-0.021}^{+0.021}$ \\[1.5mm]
\hline
\end{tabular}
\label{tab:al_wde}
\end{table}


 There is a slight tension at the 1 $\sigma$ level in the values of
 some of the cosmological parameters with respect to their values in
 the $\Lambda$CDM model. At 2 $\sigma$, however, the results are
 consistent. Indeed, when compared to the $\Lambda$CDM model, we find
 that the value of $\Theta$ is lower and the age of the universe is
 higher for all of the datasets. The value of $n_s$ is higher for the
 CMB and the CMB+CMASS datasets, and $\sigma_{\rm 8}$ is higher for
 the full dataset.

\begin{figure}
\centerline{\includegraphics[angle=-90,width=0.6\textwidth]{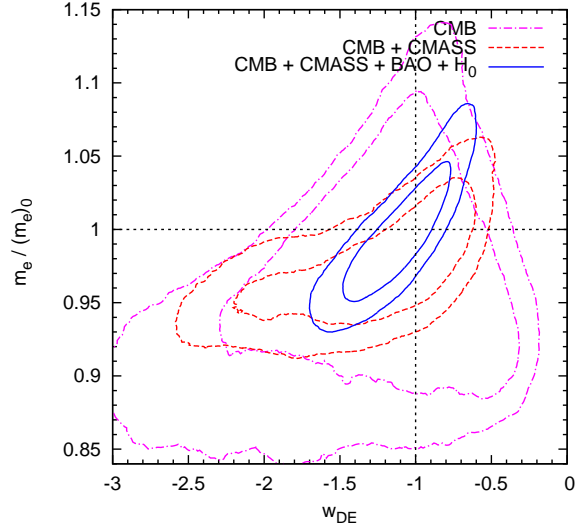}}
\caption{ The marginalized posterior distribution in the $m_e$ -
  $w_{\rm DE}$ plane for the $\Lambda$CDM parameter set extended to
  include the variation of $m_e$ and the redshift-independent value of
  $w_{\rm DE}$ as an additional parameter. The dot-dashed lines show
  the 68\% and 95\% contours obtained using CMB information alone. The
  dashed lines correspond to the results obtained from the combination
  of CMB data plus the shape of the CMASS $\xi(s)$. The solid lines
  indicate the results obtained from the full dataset combination
  (CMB+CMASS+BAO+$H_0$). The vertical and horizontal dotted lines
  correspond to the $\Lambda$CDM model, with $m_e/(m_e)_0=1$ and
  $w_{\rm DE}= -1$.}
\label{fig:2D_me_wde_variation_me-wde}
\end{figure}

 We then perform a statistical analysis varying $m_e$ and $w_{\rm
   DE}$.  Fig.~\ref{fig:2D_me_wde_variation_me-wde} presents the
 resulting constraints on the $m_e$ - $w_{\rm DE}$ plane. CMB data
 alone cannot place strong constraints on the value of $m_e$ and $w_{\rm
   DE}$. The inclusion of the CMASS correlation function restricts the
 allowed region in the parameter space, reducing the allowed range of
 values of $m_e$ by a factor of two. Using the full dataset, our bounds
 are $m_e/(m_e)_0 = 0.996 \pm 0.029$ and $w_{\rm DE}= -1.12 \pm
 0.23$. In Table~\ref{tab:me_wde} we also present the constraints on
 the other cosmological parameters.  The standard case of $m_e =
 (m_e)_0$ and $w_{\rm DE}=-1$ is completely consistent with all of the
 datasets.  The values of the cosmological parameters are consistent
 within 1 $\sigma$ with their values in the $\Lambda$CDM model.

\begin{table} 
\centering
  \caption{ The marginalized 68\% allowed regions on the cosmological
    parameters of the $\Lambda$CDM model, adding the variation of the
    electron mass $m_e$, and the dark energy equation of state $w_{\rm
      DE}$, obtained using different combinations of the datasets.  }
    \begin{tabular}{@{}lccc@{}}
    \hline
& \multirow{2}{*}{CMB}  & \multirow{2}{*}{CMB + CMASS} &  CMB + CMASS \\
&                       &                           &   + BAO + $H_0$   \\  
\hline
$w_{\rm DE}$  &  $-1.31_{-0.64}^{+0.59}$ &  $-1.30_{-0.56}^{+0.48}$ &  $-1.12_{   -0.23}^{+    0.23}$ \\[1.5mm]
$m_e / (m_e)_0$  &  $    0.962_{   -0.061}^{+    0.060}$ &  $    0.974_{   -0.029}^{+    0.029}$ &  $    0.996_{   -0.029}^{+    0.029}$ \\[1.5mm]
100$\Theta$  &  $1.012_{-0.045}^{+0.045}$ &  $1.022_{-0.021}^{+0.022}$ &  $    1.037_{-0.021}^{+    0.021}$ \\[1.5mm]
100$\omega_{\rm b}$  &  $2.13_{-0.15}^{+0.15}$ &  $2.156_{-0.076}^{+0.078}$ &  $2.204_{-0.075}^{+0.076}$ \\[1.5mm]
100$\Omega_{\rm dm}$  &  $10.78_{-0.81}^{+0.82}$ &  $10.90_{-0.64}^{+0.64}$ &  $11.52_{-0.62}^{+0.62}$ \\[1.5mm]
$\tau$  &  $0.0845_{-0.0073}^{+0.0065}$ &  $0.0849_{-0.0070}^{+0.0067}$ &  $0.0816_{-0.0068}^{+0.0060}$ \\[1.5mm]
$n_{\rm s}$  &  $0.963_{-0.012}^{+0.012}$ &  $0.963_{-0.010}^{+0.011}$ &  $0.9613_{-0.0100}^{+0.0099}$ \\[1.5mm]
ln$(10^{10}A_{\rm s})$  &  $3.077_{-0.031}^{+0.031}$ &  $3.078_{-0.031}^{+0.030}$ &  $3.087_{-0.029}^{+0.029}$ \\[1.5mm]
$\Omega_{\rm DE}$  &  $0.69_{-0.17}^{+0.15}$ &  $0.743_{-0.063}^{+0.073}$ &  $0.729_{-0.019}^{+0.019}$ \\[1.5mm]
$\Omega_{\rm m}$  &  $0.31_{-0.15}^{+0.17}$ &  $0.257_{-0.073}^{+0.063}$ &  $0.271_{-0.019}^{+0.019}$ \\[1.5mm]
$\sigma_{\rm 8}$  &  $0.83_{-0.15}^{+0.15}$ &  $0.85_{-0.11}^{+0.12}$ &  $0.852_{-0.048}^{+0.049}$ \\[1.5mm]
$t_{\rm 0}$/Gyr  &  $14.8_{-1.6}^{+1.6}$ &  $14.30_{-0.56}^{+0.56}$ &  $13.86_{-0.56}^{+0.55}$ \\[1.5mm]
$z_{\rm re}$  &  $9.9_{-1.3}^{+1.3}$ &  $   10.1_{-1.2}^{+1.2}$ &  $10.2_{-1.2}^{+1.2}$ \\[1.5mm]
$h$   &  $0.70_{-0.19}^{+0.20}$ &  $   0.730_{-0.091}^{+0.11}$ &  $0.712_{-0.020}^{+0.021}$ \\[1.5mm]
\hline
\end{tabular}
\label{tab:me_wde}
\end{table}


\subsection{Variation of fundamental constants and $f_\nu$ }
\label{Sec:alpha_fnu}

In the standard $\Lambda$CDM scenario, the dark matter component is
given entirely by cold dark matter. However, neutrino oscillations
found in recent experiments imply that they have a non-zero mass that
contributes to the total energy budget of the universe. A variation in
the neutrino mass can alter the redshift of matter-radiation equality,
thus modifying the CMB power spectrum. Furthermore, until they become
non-relativistic, neutrinos free-stream out of density perturbations,
suppressing the growth of structures on scales smaller than the
horizon at that time, which depends on their mass, thus affecting the
shape of the matter power spectrum and the correlation function.  In
this Section we explore the constraints on the neutrino fraction,
$f_\nu$, when one of the fundamental constants ($\alpha$ or $m_e$) is
allowed to vary.  We assume three neutrino species of equal mass.

\begin{figure}
\centerline{\includegraphics[angle=-90,width=0.6\textwidth]{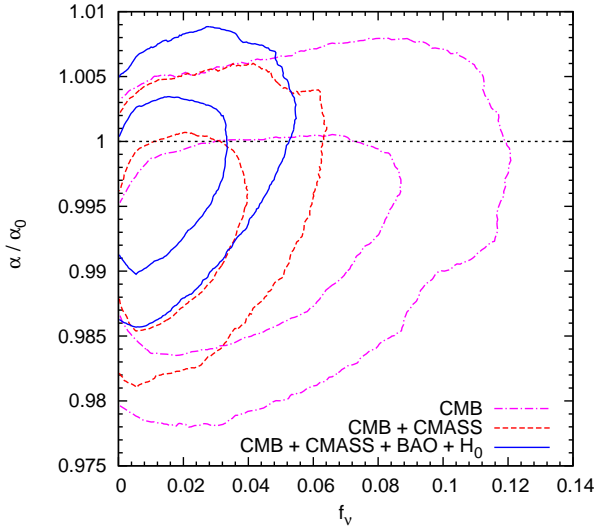}}
\caption{ The marginalized posterior distribution in the $\alpha$ -
  $f_\nu$ plane for the $\Lambda$CDM parameter set extended to include
  the variation of $\alpha$ and a non-negligible fraction of massive
  neutrinos. The dot-dashed lines show the 68\% and 95\% contours
  obtained using CMB information alone. The dashed lines correspond to
  the results obtained from the combination of CMB data plus the shape
  of the CMASS $\xi(s)$. The solid lines indicate the results obtained
  from the full dataset combination (CMB+CMASS+BAO+$H_0$). }
\label{fig:2D_al_fnu_variation_al-fnu}
\end{figure}

Fig.~\ref{fig:2D_al_fnu_variation_al-fnu} shows the contours in the
$\alpha$ - $f_\nu$ plane. The constraints on $\alpha$ are only
slightly poorer than in the $f_\nu = 0$ case. When additional datasets
are incorporated to the fit, the bounds on $\alpha$ tighten.

In Table~\ref{tab:al_fnu} we present the constraints obtained for
$\alpha$, $f_\nu$, and the remaining cosmological parameters.  In the
case of only CMB data, we find $f_\nu < 0.098$ at 95\% CL, and
$\alpha/ \alpha_0 = 0.9933_{-0.0058}^{+0.0057}$. When we also include
the information from the CMASS correlation function, this limit is
reduced to $f_\nu < 0.049$ at 95\% CL, and the constraint on $\alpha$
is $\alpha/ \alpha_0 = 0.9940_{-0.0049}^{+0.0050}$.  The constraints
for the full data set are $f_\nu < 0.043$ at 95\% CL and $\alpha/
\alpha_0 = 0.9978_{-0.0045}^{+0.0044}$. We find no degeneracy between
$\alpha$ and $f_\nu$. When all datasets are included in the analysis,
the value of $\alpha$ is increased (due to the larger value of $h$),
our results being consistent with no variation of $\alpha$ within 1
$\sigma$.

\begin{table*} 
\centering
  \caption{ The marginalized 68\% allowed regions on the cosmological
    parameters of the $\Lambda$CDM model, adding the variation of the
    fine structure constant $\alpha$ and $f_\nu$ as a free parameter,
    obtained using different combinations of the datasets.  }
    \begin{tabular}{@{}lccc@{}}
    \hline
& \multirow{2}{*}{CMB}  & \multirow{2}{*}{CMB + CMASS} &  CMB + CMASS \\
&                       &                           &   + BAO + $H_0$   \\  
\hline
 $f_\nu$  &  $< 0.098$ (95\% CL) &  $< 0.049$ (95\% CL) &  $< 0.043$ (95\% CL)  \\[1.5mm]
 $\alpha/ \alpha_0$  & $0.9933_{-0.0058}^{+0.0057}$ &  $0.9940_{-0.0049}^{+0.0050}$ &  $0.9978_{-0.0045}^{+0.0044}$ \\[1.5mm]
100$\Theta$  &  $1.0312_{-0.0082}^{+0.0081}$ &  $1.0326_{-0.0069}^{+0.0071}$ &  $1.0382_{-0.0062}^{+0.0061}$ \\[1.5mm]
100$\omega_{\rm b}$  &  $2.187_{-0.044}^{+0.044}$ &  $2.209_{-0.038}^{+0.039}$ &  $2.228_{-0.037}^{+0.038}$ \\[1.5mm]
100$\omega_{\rm dm}$  &  $11.86_{-0.70}^{+0.71}$ &  $11.25_{-0.39}^{+0.39}$ &  $11.27_{-0.34}^{+0.34}$ \\[1.5mm]
$\tau$  &  $0.0857_{-0.0072}^{+0.0061}$ &  $0.0889_{-0.0075}^{+0.0068}$ &  $0.0869_{-0.0070}^{+0.0063}$ \\[1.5mm]
 $n_{\rm s}$  &  $0.968_{-0.016}^{+0.016}$ &  $0.974_{-0.013}^{+0.013}$ &  $0.972_{-0.012}^{+0.013}$ \\[1.5mm]
 ln$(10^{10}A_{\rm s})$  &  $3.097_{-0.034}^{+0.034}$ &  $3.094_{-0.032}^{+0.033}$ &  $3.086_{-0.031}^{+0.031}$ \\[1.5mm]
$\sum m_\nu$  &  $< 1.2\,{\rm eV}$ (95\% CL) & $< 0.53\,{\rm eV}$ (95\% CL)  & $< 0.46\,{\rm eV}$ (95\% CL) \\[1.5mm] 
 $\Omega_{\rm DE}$  &  $0.636_{-0.062}^{+0.062}$ &  $0.696_{-0.020}^{+0.020}$ &  $0.714_{-0.011}^{+0.011}$ \\[1.5mm]
$\Omega_{\rm m}$  &  $0.364_{-0.062}^{+0.062}$ &  $0.304_{-0.020}^{+0.020}$ &  $0.286_{-0.011}^{+0.011}$ \\[1.5mm]
$\sigma_{\rm 8}$  &  $0.715_{-0.074}^{+0.073}$ &  $0.761_{-0.044}^{+0.043}$ &  $0.772_{-0.039}^{+ 0.038}$ \\[1.5mm]
 $t_{\rm 0}$/Gyr  &  $14.32_{-0.28}^{+0.28}$ &  $14.11_{-0.20}^{+0.20}$ &  $13.90_{-0.16}^{+0.16}$ \\[1.5mm]
$z_{\rm re}$  &  $10.8_{-1.2}^{+1.3}$ &  $10.9_{-1.3}^{+1.3}$ &  $10.6_{-1.2}^{+1.2}$ \\[1.5mm]
$h$  &  $   0.626_{-0.041}^{+0.041}$ &  $0.666_{-0.019}^{+0.019}$ &  $0.688_{-0.012}^{+0.012}$ \\[1.5mm]
\hline
\end{tabular}
\label{tab:al_fnu}
\end{table*}


Regarding the constraints on the primary cosmological parameters, we
find that they are consistent with the $\Lambda$CDM values at 1
$\sigma$ in most of the cases.  For the CMB and CMB+CMASS datasets,
some of the derived parameters differ by more than 1 $\sigma$ from
their values in the $\Lambda$CDM model, but they are consistent within
1 $\sigma$ when we consider the full dataset, with exception of
$\sigma_{\rm 8}$, which is lower by more than 1 $\sigma$ also for the
full dataset. For the CMB dataset, $h$ differs by more than 2 $\sigma$
from the $\Lambda$CDM value, by 1 $\sigma$ for the CMB+CMASS dataset,
and is fully consistent for the full dataset. The constraints at 95\%
CL on the sum of the neutrinos masses are $\sum m_\nu < 1.2\,{\rm eV}$
(CMB), $\sum m_\nu < 0.53\,{\rm eV}$ (CMB+CMASS), and $\sum m_\nu <
0.46\,{\rm eV}$ (full dataset).

\begin{figure}
\centerline{\includegraphics[angle=-90,width=0.6\textwidth]{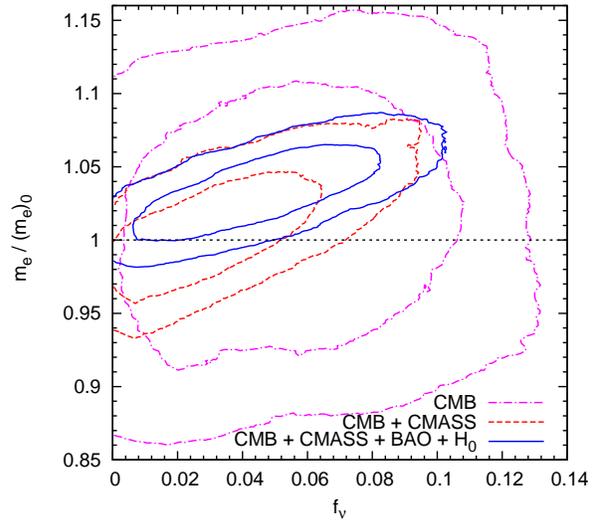}}
\caption{ The marginalized posterior distribution in the $m_e$ -
  $f_\nu$ plane for the $\Lambda$CDM parameter set extended to include
  the variation of $m_e$ and a non-negligible fraction of massive
  neutrinos. The dot-dashed lines show the 68\% and 95\% contours
  obtained using CMB information alone. The dashed lines correspond to
  the results obtained from the combination of CMB data plus the shape
  of the CMASS $\xi(s)$. The solid lines indicate the results obtained
  from the full dataset combination (CMB+CMASS+BAO+$H_0$). }
\label{fig:2D_me_fnu_variation_me-fnu}
\end{figure}

Fig.~\ref{fig:2D_me_fnu_variation_me-fnu} shows the contours for the
analysis in the $m_e$ - $f_\nu$ plane. The precision in the
constraints on $m_e$ are similar as those in the $f_\nu = 0$
case. When the CMASS correlation function is added to the CMB dataset,
the allowed region in the parameter space is reduced. Due to the
degeneracy between $H_0$ and $m_e$, and between the latter and $f_\nu$
(correlation factor equal to 0.83 and 0.72, respectively), when the
full dataset is considered, the contours are shifted towards a region
with higher values of $m_e$ and $f_\nu$.

In Table~\ref{tab:me_fnu} we present the constraints obtained for
$m_e$, $f_\nu$, and the cosmological parameters.  In the case of only
CMB data, we find $f_\nu < 0.11$ at 95\% CL, and $m_e/(m_e)_0 =
1.011_{-0.066}^{+0.067}$. When we also include the information from
the CMASS correlation function, this limit is reduced to $f_\nu <
0.074$ at 95\% CL, and the constraint on $m_e$ is $m_e/(m_e)_0 =
1.009_{-0.028}^{+0.029}$.  The constraints for the full data set are
$f_\nu < 0.086$ at 95\% CL and $m_e/(m_e)_0 = 1.035 \pm 0.021$. When
all datasets are included in the analysis, the value of $m_e$ is
increased (due to the larger value of $h$), and our results are
consistent with no variation of $m_e$ at 2 $\sigma$.

\begin{table*} 
\centering
  \caption{ The marginalized 68\% allowed regions on the cosmological
    parameters of the $\Lambda$CDM model, adding the variation of the
    electron mass $m_e$ and $f_\nu$ as a free parameter, obtained
    using different combinations of the datasets.  }
    \begin{tabular}{@{}lccc@{}}
    \hline
& \multirow{2}{*}{CMB}  & \multirow{2}{*}{CMB + CMASS} &  CMB + CMASS \\
&                       &                           &   + BAO + $H_0$   \\  
\hline
$f_\nu $  & $ < 0.11 (95\% {\rm CL})$  &   $< 0.074 (95\% {\rm CL})$ & $< 0.086 (95\% {\rm CL})$  \\[1.5mm]
$m_e/(m_e)_0$  &  $1.011_{-0.066}^{+0.067}$ &  $1.009_{-0.028}^{+0.029}$ &  $1.035_{-0.021}^{+0.021}$ \\[1.5mm]
100$\Theta$  &  $1.048_{-0.047}^{+0.048}$ &  $1.047_{-0.020}^{+0.021}$ &  $1.065_{-0.015}^{+0.014}$ \\[1.5mm]
100$\omega_{\rm b}$  &  $2.22_{-0.16}^{+0.16}$ &  $2.225_{-0.064}^{+0.065}$ &  $2.276_{-0.048}^{+0.047}$ \\[1.5mm]
100$\omega_{\rm dm}$  &  $12.2_{-1.1}^{+1.1}$ &  $11.81_{-0.89}^{+0.92}$ &  $12.60_{-0.76}^{+0.77}$ \\[1.5mm]
$\tau $  &  $0.0840_{-0.0073}^{+0.0063}$ &  $0.0839_{-0.0072}^{+0.0061}$ &  $0.0818_{-0.0067}^{+0.0061}$ \\[1.5mm]
 $n_{\rm s}$  &  $0.957_{-0.014}^{+0.013}$ &  $0.962_{-0.010}^{+0.010}$ &  $0.9593_{-0.0099}^{+0.0098}$ \\[1.5mm]
ln$(10^{10}A_{\rm s})$  &  $3.081_{-0.031}^{+0.030}$ &  $3.081_{-0.031}^{+0.031}$ &  $3.089_{-0.030}^{+0.029}$ \\[1.5mm]
$\sum m_\nu $  & $ < 1.4\,{\rm eV}$ (95\% CL)  &   $ < 0.91\,{\rm eV}$ (95\% CL) & $< 1.1\,{\rm eV}$ (95\% CL)  \\[1.5mm]
 $\Omega_{\rm DE}$  &  $0.64_{-0.14}^{+0.13}$ &  $0.702_{-0.021}^{+0.021}$ &  $0.711_{-0.011}^{+0.012}$ \\[1.5mm]
 $\Omega_{\rm m}$  &  $0.36_{-0.13}^{+0.14}$ &  $0.297_{-0.021}^{+0.021}$ &  $0.289_{-0.012}^{+0.011}$ \\[1.5mm]
$\sigma_{\rm 8}$  &  $0.694_{-0.091}^{+0.090}$ &  $0.745_{-0.050}^{+0.048}$ &  $0.748_{-0.055}^{+0.054}$ \\[1.5mm]
 $t_{\rm 0}$/Gyr  &  $14.0_{-1.5}^{+1.6}$ &  $13.76_{-0.54}^{+0.52}$ &  $13.28_{-0.33}^{+0.33}$ \\[1.5mm]
$z_{\rm re}$  &  $10.7_{-1.4}^{+1.4}$ &  $10.6_{-1.3}^{+1.3}$ &  $10.8_{-1.3}^{+1.3}$ \\[1.5mm]
$h$  &  $   0.67_{-0.15}^{+0.15}$ &  $0.689_{-0.035}^{+0.037}$ &  $0.718_{-0.019}^{+0.019}$ \\[1.5mm]
\hline
\end{tabular}
\label{tab:me_fnu}
\end{table*}


With regards to the constraints on the cosmological parameters, we
find that they are consistent with the $\Lambda$CDM values at 1
$\sigma$ in most of the cases, although $\omega_{\rm dm}$ differs from
the $\Lambda$CDM value by 1 $\sigma$ for the full dataset.  There is
also tension for $\sigma_{\rm 8}$ at the 1 $\sigma$ level for all of
the datasets, in the sense that our constraints are lower than the
$\Lambda$CDM values. For the full dataset, the age of the universe is
lower and $h$ is higher by more than 1 $\sigma$ than their
corresponding values in the $\Lambda$CDM model. The constraints at
95\% CL on the sum of the neutrinos masses are $\sum m_\nu < 1.4\,{\rm
  eV}$ (CMB), $\sum m_\nu < 0.91\,{\rm eV}$ (CMB+CMASS), and $\sum
m_\nu < 1.1\,{\rm eV}$ (full dataset).

\subsection{Variation of fundamental constants and $N_{\rm eff}$ }
\label{Sec:alpha_Neff}

The effective number of relativistic species, $N_{\rm eff}$, has been
reported as higher than the expected standard value of $N_{\rm eff} =
3.046$ \citep{Dunkley2011,Keisler2011}. This may indicate either
additional relativistic species, or evidence for non-standard
decoupling.
Here we study the constraints from the CMASS correlation function on
the variation of $\alpha$ and $m_e$ when $N_{\rm eff}$ can differ from
its standard value, providing details on the constraints set by the
different datasets considered in this paper.

In the case of no variation of the fundamental constants,
\citet{Keisler2011} present a bound of $N_{\rm eff} = 3.85 \pm 0.62$
for the CMB dataset (WMAP+SPT). We consider the $\Lambda$CDM+$N_{\rm
  eff}$ model and obtain the constraints $N_{\rm eff} = 3.75 \pm 0.58$
for CMB+CMASS and $N_{\rm eff} = 3.86_{-0.40}^{+0.39}$ for the full
dataset.

\begin{figure}
\centerline{\includegraphics[angle=-90,width=0.6\textwidth]{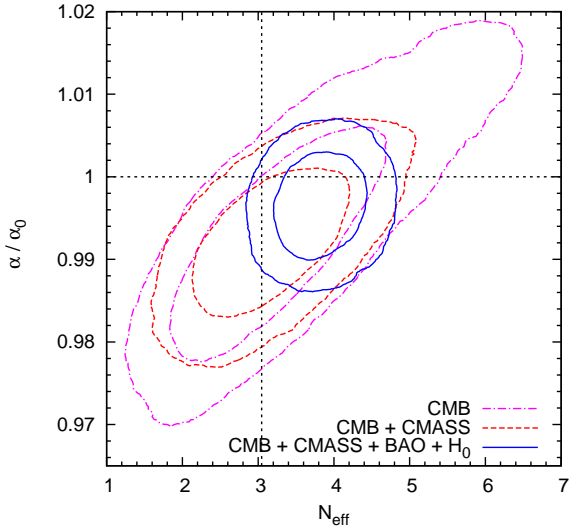}}
\caption{ The marginalized posterior distribution in the $\alpha$ -
  $N_{\rm eff}$ plane for the $\Lambda$CDM parameter set extended to
  include the variation of $\alpha$ and a variable effective number of
  relativistic degrees of freedom $N_{\rm eff}$.  The dot-dashed lines
  show the 68\% and 95\% contours obtained using CMB information
  alone. The dashed lines correspond to the results obtained from the
  combination of CMB data plus the shape of the CMASS $\xi(s)$. The
  solid lines indicate the results obtained from the full dataset
  combination (CMB+CMASS+BAO+$H_0$). }
\label{fig:2D_al_Nnu_variation_al-Nnu}
\end{figure}

The contours in Fig.~\ref{fig:2D_al_Nnu_variation_al-Nnu} show the
two-dimensional marginalized constraints in the $N_{\rm eff}$ -
$\alpha$ plane.  There is a strong degeneracy between $\alpha$ and
$N_{\rm eff}$ in the CMB dataset. When we add the CMASS dataset, this
degeneracy is partially alleviated, and it disappears for the full
dataset. The inclusion of the HST $H_0$ prior shifts the contours
towards higher values of $N_{\rm eff}$ and $\alpha$, making the latter
consistent with its present value at 1 $\sigma$.

In Table~\ref{tab:al_Nnu} we present the constraints on $N_{\rm eff}$,
$\alpha$, and the rest of the cosmological parameters for all of our
datasets.  When $\alpha$ is allowed to vary, $N_{\rm eff} =
3.37_{-0.97}^{+0.95}$ for the CMB dataset, which is a value more
consistent with the standard one than when $\alpha$ is fixed to its
present value \citep[as said before, for WMAP+SPT data, the value
  presented by][is $N_{\rm eff} = 3.85 \pm 0.62$]{Keisler2011}.  The
constraint on $\alpha$ is $\alpha/\alpha_0 = 0.9935 \pm 0.0093$. By
adding the CMASS correlation function, we tighten the bounds on
$N_{\rm eff}$ and $\alpha$, which become $N_{\rm eff}=
3.21_{-0.67}^{+0.66}$ and $\alpha/\alpha_0 = 0.9926 \pm 0.0058$.  For
CMB+CMASS+BAO, $N_{\rm eff} = 3.19 \pm 0.64$ and $\alpha/\alpha_0 =
0.9926_{-0.0055}^{+0.0054}$.  When all datasets are considered in the
analysis, the constraints are $N_{\rm eff}= 3.83 \pm 0.40$ and
$\alpha/\alpha_0 = 0.9967 \pm 0.0042$. The $H_0$ prior shifts $N_{\rm
  eff}$ towards higher values, which lie almost 2 $\sigma$ from the
standard value of 3.046. On the other hand, $\alpha$ is consistent
with no variation at 1 $\sigma$. \citet{Menegoni2012} studied the
variation of $\alpha$ when $N_{\rm eff}$ can have values different
from the standard one, allowing also for variations in the primordial
helium abundance, $Y_{\rm He}$.  Their constraints are
$\alpha/\alpha_0 = 0.990 \pm 0.006$ and $N_{\rm eff} =
4.10_{-0.29}^{+0.24}$. The difference in the datasets is that they use
the DR7 LRG power spectrum and add the ACT data to their analysis,
presenting their results only for the full dataset.

\begin{table} 
\centering
  \caption{ The marginalized 68\% allowed regions on the cosmological
    parameters of the $\Lambda$CDM model, adding the variation of the
    fine structure constant $\alpha$ and $N_{\rm eff}$ as a free
    parameter, obtained using different combinations of the
    datasets. }
\begin{tabular}{@{}lccc@{}}
\hline
& \multirow{2}{*}{CMB}  & \multirow{2}{*}{CMB + CMASS} &  CMB + CMASS \\
&                       &                           &   + BAO + $H_0$   \\  
\hline
$N_{\rm eff}$  &  $3.37_{-0.97}^{+0.95}$ &  $3.21_{-0.67}^{+0.66}$ &  $3.83_{-0.40}^{+0.40}$ \\[1.5mm]
$\alpha/ \alpha_0$  &  $0.9935_{-0.0093}^{+0.0093}$ &  $0.9926_{-0.0058}^{+0.0058}$ &  $0.9967_{-0.0042}^{+0.0042}$ \\[1.5mm]
100$\Theta$  &  $1.032_{-0.012}^{+0.012}$ &  $1.0304_{-0.0074}^{+0.0075}$ &  $1.0349_{-0.0058}^{+0.0057}$ \\[1.5mm]
100$\omega_{\rm b}$  &  $2.228_{-0.079}^{+0.079}$ &  $2.218_{-0.055}^{+0.055}$ &  $2.254_{-0.041}^{+0.041}$ \\[1.5mm]
100$\omega_{\rm dm}$  &  $11.7_{-1.8}^{+1.7}$ &  $11.4_{-1.4}^{+1.3}$ &  $12.72_{-0.87}^{+0.88}$ \\[1.5mm]
$\tau$  &  $0.0886_{-0.0074}^{+0.0063}$ &  $0.0882_{-0.0073}^{+0.0064}$ &  $0.0878_{-0.0072}^{+0.0065}$ \\[1.5mm]
$n_{\rm s}$  &  $0.981_{-0.021}^{+0.021}$ &  $0.979_{-0.017}^{+0.017}$ &  $0.988_{-0.015}^{+0.015}$ \\[1.5mm]
ln$(10^{10}A_{\rm s})$  &  $3.112_{-0.049}^{+0.049}$ &  $3.106_{-0.045}^{+0.044}$ &  $3.134_{-0.036}^{+0.036}$ \\[1.5mm]
$\Omega_{\rm DE}$  &  $0.705_{-0.049}^{+0.050}$ &  $0.707_{-0.019}^{+0.019}$ &  $0.715_{-0.010}^{+0.010}$ \\[1.5mm]
$\Omega_{\rm m}$  &  $0.295_{-0.050}^{+0.049}$ &  $0.293_{-0.019}^{+0.019}$ &  $0.285_{-0.010}^{+0.010}$ \\[1.5mm]
$\sigma_{\rm 8}$  &  $0.832_{-0.061}^{+0.060}$ &  $0.822_{-0.049}^{+0.048}$ &  $0.867_{-0.032}^{+0.032}$ \\[1.5mm]
 $t_{\rm 0}$/Gyr  &  $13.8_{-1.3}^{+1.3}$ &  $13.95_{-0.79}^{+0.81}$ &  $13.20_{-0.38}^{+0.38}$ \\[1.5mm]
$z_{\rm re}$  &  $10.9_{-1.3}^{+1.3}$ &  $10.9_{-1.3}^{+1.3}$ &  $11.1_{-1.3}^{+1.3}$ \\[1.5mm]
$h$  &  $0.696_{-0.092}^{+0.092}$ &  $0.683_{-0.048}^{+0.047}$ &  $0.725_{-0.022}^{+0.022}$ \\[1.5mm]
\hline
\end{tabular}
\label{tab:al_Nnu}
\end{table}


There is tension at the 1 $\sigma$ level in some of the cosmological
parameters with respect to their $\Lambda$CDM model values. For the
full dataset, $\Theta$ and $t_{\rm 0}$ are more than 1 $\sigma$ lower,
and $n_{\rm s}$, $\sigma_{\rm 8}$, $h$ and $\omega_{\rm dm}$ are
larger by more than 1 $\sigma$. $\Theta$ is also more than 1 $\sigma$
lower than its concordance value for the CMB+CMASS dataset. In the
rest of the cases, the parameters are all consistent within 1 $\sigma$
with their $\Lambda$CDM values.

\begin{figure}
\centerline{\includegraphics[angle=-90,width=0.6\textwidth]{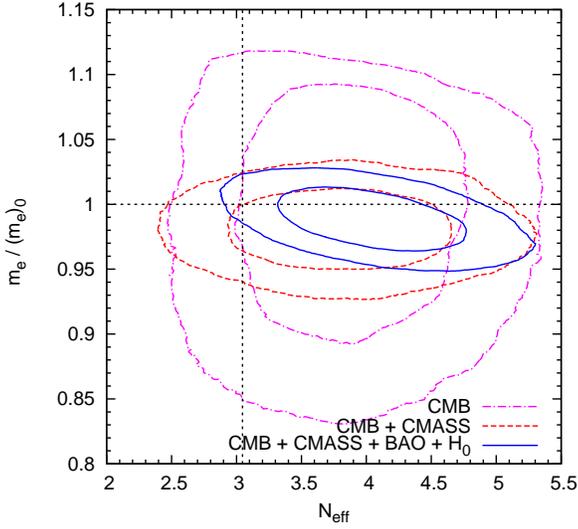}}
\caption{ The marginalized posterior distribution in the $m_e$ -
  $N_{\rm eff}$ plane for the $\Lambda$CDM parameter set extended to
  include the variation of $m_e$ and a variable effective number of
  relativistic degrees of freedom $N_{\rm eff}$.  The dot-dashed lines
  show the 68\% and 95\% contours obtained using CMB information
  alone. The dashed lines correspond to the results obtained from the
  combination of CMB data plus the shape of the CMASS $\xi(s)$. The
  solid lines indicate the results obtained from the full dataset
  combination (CMB+CMASS+BAO+$H_0$). }
\label{fig:2D_me_Nnu_variation_me-Nnu}
\end{figure}

We now analyze the results for the case in which $m_e$ and $N_{\rm
  eff}$ are allowed to vary. In
Fig.~\ref{fig:2D_me_Nnu_variation_me-Nnu} we show the two-dimensional
marginalized constraints in the $N_{\rm eff}$ - $m_e$ plane. There is
no degeneracy between $m_e$ and $N_{\rm eff}$ for the CMB and
CMB+CMASS datasets. When the full dataset is considered, a mild
correlation between both quantities appears.

\begin{table} 
\centering
  \caption{ The marginalized 68\% allowed regions on the cosmological
    parameters of the $\Lambda$CDM model, adding the variation of the
    electron mass $m_e$ and $N_{\rm eff}$ as a free
    parameter, obtained using different combinations of the
    datasets. }
\begin{tabular}{@{}lccc@{}}
\hline
& \multirow{2}{*}{CMB}  & \multirow{2}{*}{CMB + CMASS} &  CMB + CMASS \\
&                       &                           &   + BAO + $H_0$   \\  
\hline
$N_{\rm eff}$  &  $3.86_{-0.60}^{+0.60}$ &  $3.80_{-0.57}^{+0.57}$ &  $4.04_{-0.46}^{+0.47}$ \\[1.5mm]
$m_e /(m_e)_0$  &  $0.988_{-0.065}^{+0.065}$ &  $0.981_{-0.020}^{+0.020}$ &  $0.988_{-0.016}^{+0.015}$ \\[1.5mm]
100$\Theta$  &  $1.030_{-0.048}^{+0.048}$ &  $1.026_{-0.015}^{+0.015}$ &  $1.031_{-0.011}^{+0.011}$ \\[1.5mm]
100$\omega_{\rm b}$  &  $2.24_{-0.17}^{+0.17}$ &  $2.216_{-0.062}^{+0.062}$ &  $2.245_{-0.042}^{+0.042}$ \\[1.5mm]
100$\omega_{\rm dm}$  &  $12.4_{-1.4}^{+1.5}$ &  $12.2_{-1.2}^{+1.2}$ &  $12.85_{-0.80}^{+0.82}$ \\[1.5mm]
$\tau$  &  $0.0891_{-0.0075}^{+0.0067}$ &  $0.0884_{-0.0075}^{+0.0065}$ &  $0.0882_{-0.0075}^{+0.0066}$ \\[1.5mm]
$n_{\rm s}$  &  $0.986_{-0.020}^{+0.019}$ &  $0.985_{-0.017}^{+0.017}$ &  $0.990_{-0.016}^{+0.016}$ \\[1.5mm]
ln$(10^{10}A_{\rm s})$  &  $3.124_{-0.045}^{+0.044}$ &  $3.119_{-0.042}^{+0.042}$ &  $3.134_{-0.036}^{+0.036}$ \\[1.5mm]
$\Omega_{\rm DE}$  &  $0.69_{-0.11}^{+0.11}$ &  $0.709_{-0.020}^{+0.020}$ &  $0.715_{-0.011}^{+0.011}$ \\[1.5mm]
$\Omega_{\rm m}$  &  $0.31_{-0.11}^{+0.11}$ &  $0.291_{-0.020}^{+0.020}$ &  $0.285_{-0.011}^{+0.011}$ \\[1.5mm]
$\sigma_{\rm 8}$  &  $0.844_{-0.086}^{+0.085}$ &  $0.835_{-0.049}^{+0.049}$ &  $0.862_{-0.033}^{+0.033}$ \\[1.5mm]
$t_{\rm 0}$/Gyr  &  $13.4_{-1.6}^{+1.6}$ &  $13.53_{-0.65}^{+0.65}$ &  $13.14_{-0.34}^{+0.34}$ \\[1.5mm]
$z_{\rm re}$  &  $10.9_{-1.5}^{+1.5}$ &  $10.7_{-1.3}^{+1.3}$ &  $10.9_{-1.3}^{+1.3}$ \\[1.5mm]
$h$  &  $0.73_{-0.16}^{+0.16}$ &  $0.705_{-0.043}^{+0.042}$ &  $0.728_{-0.020}^{+0.020}$ \\[1.5mm]
\hline
\end{tabular}
\label{tab:me_Nnu}
\end{table}


In Table~\ref{tab:me_Nnu} we present the constraints on $N_{\rm eff}$,
$m_e$, and the rest of the cosmological parameters for all of our
datasets.  For the CMB dataset, $N_{\rm eff} = 3.86 \pm 0.60$, which
is more than 1 $\sigma$ higher than the standard value.  The
constraint on $m_e$ is $m_e/(m_e)_0 = 0.988 \pm 0.065$. By adding the
CMASS correlation function, we significantly tighten the bounds on
$m_e$ but only slightly improve the limits on $N_{\rm eff}$, which
become $N_{\rm eff}= 3.80 \pm 0.57$ and $m_e /(m_e)_0 = 0.981 \pm
0.020$. For CMB+CMASS+BAO, $N_{\rm eff}= 3.78 \pm 0.58$ and $m_e
/(m_e)_0 = 0.981 \pm 0.018$.  When all datasets are considered in the
analysis, the constraints are $N_{\rm eff}= 4.04_{-0.46}^{+0.47}$ and
$m_e /(m_e)_0 = 0.988_{-0.016}^{+0.015}$. The $H_0$ prior shifts
$N_{\rm eff}$ towards even higher values, which are more than 2
$\sigma$ above the standard value of 3.046. On the other hand, $m_e$
is consistent with no variation at the 1 $\sigma$ level for all of the
datasets.

 Regarding the values of the remaining cosmological parameters, they
 are all consistent within 1 $\sigma$ with their $\Lambda$CDM values
 for the CMB dataset. For the CMB+CMASS dataset, $n_{\rm s}$ is higher
 than its standard value by more than 1 $\sigma$. For the full
 dataset, $\omega_{\rm dm}$, $n_{\rm s}$, $\sigma_{\rm 8}$, and $h$
 are higher than their standard values in more than 1 $\sigma$, while
 $t_{\rm 0}$ is lower in more than 1 $\sigma$.

\subsection{Variation of fundamental constants and $Y_{\rm He}$ }
\label{Sec:alpha_YHe}

Light nuclei begin to form in a process known as big bang
nucleosynthesis
\citep[BBN,][]{Alpher1948,SchrammTurner1998,Steigman2007}, when the
universe cools to $T \sim 0.1 \, {\rm MeV}$.  We denote the primordial
abundance (mass fraction) of ${}^4$He as $Y_{\rm He}$, which is a
function of the baryon density and the expansion rate during BBN. The
value of $Y_{\rm He}$ can be estimated by the effect of helium on the
CMB damping tail.  Helium combines earlier than hydrogen, and thus
more helium (at fixed baryon density) leads to fewer free electrons
during hydrogen recombination. This, in turn, leads to larger
diffusion lengths for photons and less power in the CMB damping tail.
We study the constraints from the CMASS correlation function on the
variation of $\alpha$ and $m_e$ when $Y_{\rm He}$ can differ from its
standard value of $Y_{\rm He}=0.24$. 

In the case of no variation of the fundamental constants,
\citet{Keisler2011} present a bound of $Y_{\rm He} = 0.296 \pm 0.030$
for the CMB dataset (WMAP+SPT). We consider the $\Lambda$CDM+$Y_{\rm
  He}$ model and obtain the constraints $Y_{\rm He} = 0.297 \pm 0.030$
for CMB+CMASS and $Y_{\rm He} = 0.301_{-0.029}^{+0.028}$ for the full
dataset.

\begin{figure}
\centerline{\includegraphics[angle=-90,width=0.6\textwidth]{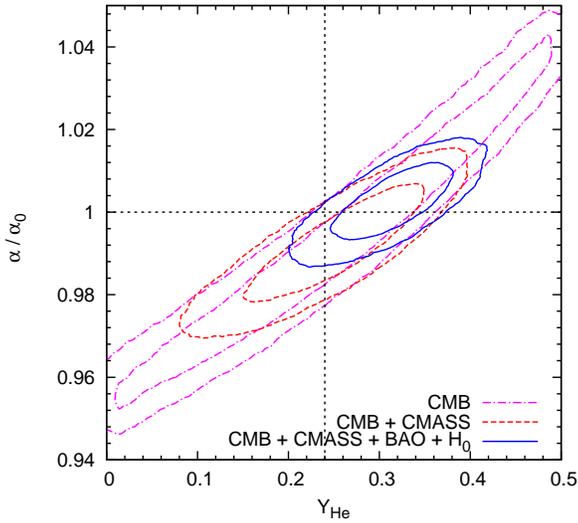}}
\caption{ The marginalized posterior distribution in the $\alpha$ -
  $Y_{\rm He}$ plane for the $\Lambda$CDM parameter set extended to
  include the variation of $\alpha$ and a variable primordial helium
  abundance $Y_{\rm He}$. The dot-dashed lines show the 68\% and 95\% contours obtained using CMB information alone. The dashed lines
  correspond to the results obtained from the combination of CMB data
  plus the shape of the CMASS $\xi(s)$. The solid lines indicate the
  results obtained from the full dataset combination
  (CMB+CMASS+BAO+$H_0$). }
\label{fig:2D_al_YHe_variation_al-YHe}
\end{figure}

The contours in Fig.~\ref{fig:2D_al_YHe_variation_al-YHe} show the
two-dimensional marginalized constraints in the $Y_{\rm He}$ -
$\alpha$ plane.  There is a strong degeneracy between $\alpha$ and
$Y_{\rm He}$ in all of the datasets. When $\alpha$ is allowed to vary,
CMB information alone is insufficient to place any reliable constraint on
$Y_{\rm He}$. When the CMASS dataset is added, the constraint improves
noticeably.  The inclusion of the HST $H_0$ prior improves further the
constraints, and shifts the contours towards higher values of $Y_{\rm
  He}$ and $\alpha$, making the former measurement inconsistent with its standard
value at 1 $\sigma$.

In Table~\ref{tab:al_YHe} we present the constraints on $Y_{\rm He}$,
$\alpha$, and the rest of the cosmological parameters for all of our
datasets.  When $\alpha$ is allowed to vary, $Y_{\rm He} = 0.26 \pm
0.15$ for the CMB dataset, which is a value consistent with the
standard one, contrary to the case when $\alpha$ is fixed to its
present value \citep[for WMAP+SPT data, the value presented by][is
  $Y_{\rm He} = 0.296 \pm 0.030$]{Keisler2011}.  The constraint on
$\alpha$ is $\alpha/\alpha_0 = 0.994_{-0.025}^{+0.026}$. By adding the
CMASS correlation function, we tighten the bounds on $Y_{\rm He}$ and
$\alpha$ to $Y_{\rm He}= 0.249 \pm 0.064$ and $\alpha/\alpha_0 =
0.9920_{-0.0093}^{+0.0092}$.  When all datasets are considered in the
analysis, the constraints are $Y_{\rm He} = 0.314 \pm 0.043$
and $\alpha/\alpha_0 = 1.0023_{-0.0061}^{+0.0062}$. The $H_0$ prior
shifts $Y_{\rm He}$ towards higher values, which are almost 2 $\sigma$
above the standard value of 0.24. On the other hand, $\alpha$ is
consistent with no variation at 1 $\sigma$. \citet{Menegoni2012}
studied the joint variation of $\alpha$, $N_{\rm eff}$ and $Y_{\rm
  He}$; their constraint is $Y_{\rm He}= 0.215 \pm 0.096$ but with
$N_{\rm eff}$ higher by more than 3 $\sigma$ than its standard value.

\begin{table} 
\centering
  \caption{
    The marginalized 68\% allowed regions on the cosmological parameters of the $\Lambda$CDM model, adding the variation of the fine structure constant $\alpha$ and $Y_{\rm He}$ as a free parameter,
    obtained using different combinations of the datasets.
}
    \begin{tabular}{@{}lccc@{}}
    \hline
& \multirow{2}{*}{CMB}  & \multirow{2}{*}{CMB + CMASS} &  CMB + CMASS \\
&                       &                           &   + BAO + $H_0$   \\  
\hline
$Y_{\rm He}$  &  $0.26_{-0.15}^{+0.15}$ &  $0.249_{-0.064}^{+0.064}$ &  $0.314_{-0.043}^{+0.043}$ \\[1.5mm]
$\alpha / \alpha_0$  &  $0.994_{-0.025}^{+0.026}$ &  $0.9920_{-0.0093}^{+0.0092}$ &  $ 1.0023_{-0.0061}^{+0.0062}$ \\[1.5mm]
100$\Theta$  &  $1.034_{-0.041}^{+0.042}$ &  $1.030_{-0.015}^{+0.015}$ &  $1.0466_{-0.0096}^{+0.0096}$ \\[1.5mm]
100$\omega_{\rm b}$  &  $2.23_{-0.14}^{+0.15}$ &  $2.214_{-0.060}^{+0.061}$ &  $2.265_{-0.046}^{+0.047}$ \\[1.5mm]
100$\omega_{\rm dm}$  &  $11.26_{-0.77}^{+0.78}$ &  $11.13_{-0.66}^{+0.66}$ &  $11.84_{-0.57}^{+0.57}$ \\[1.5mm]
$\tau$  &  $0.0881_{-0.0074}^{+0.0065}$ &  $0.0878_{-0.0073}^{+0.0062}$ &  $0.0856_{-0.0069}^{+0.0063}$ \\[1.5mm]
$n_{\rm s}$  &  $0.978_{-0.015}^{+0.014}$ &  $0.978_{-0.013}^{+0.013}$ &  $0.977_{-0.013}^{+0.013}$ \\[1.5mm]
ln$(10^{10}A_{\rm s})$  &  $3.105_{-0.036}^{+0.036}$ &  $3.102_{-0.035}^{+0.036}$ &  $3.117_{-0.033}^{+0.033}$ \\[1.5mm]
$\Omega_{\rm DE}$  &  $0.695_{-0.098}^{+0.094}$ &  $0.705_{-0.020}^{+0.020}$ &  $0.718_{-0.010}^{+0.010}$ \\[1.5mm]
$\Omega_{\rm m}$  &  $0.305_{-0.094}^{+0.098}$ &  $0.295_{-0.020}^{+0.020}$ &  $0.2817_{-0.010}^{+0.010}$ \\[1.5mm]
$\sigma_{\rm 8}$  &  $0.822_{-0.068}^{+0.068}$ &  $0.816_{-0.044}^{+0.044}$ &  $0.862_{-0.036}^{+0.036}$ \\[1.5mm]
$t_{\rm 0}$/Gyr &  $14.0_{-1.4}^{+1.4}$ &  $14.08_{-0.47}^{+0.47}$ &  $13.56_{-0.28}^{+0.28}$ \\[1.5mm]
$z_{\rm re}$  &  $10.9_{-1.3}^{+1.3}$ &  $10.8_{-1.3}^{+1.2}$ &  $10.9_{-1.3}^{+1.3}$ \\[1.5mm]
$h$  &  $0.69_{-0.13}^{+0.13}$ &  $0.674_{-0.033}^{+0.032}$ &  $0.708_{-0.017}^{+0.017}$ \\[1.5mm]
\hline
\end{tabular}
\label{tab:al_YHe}
\end{table}


There is a slight tension at the 1 $\sigma$ level for the parameter
$n_{\rm s}$, which is higher than its $\Lambda$CDM value for the
CMB+CMASS and the full datasets.  In the rest of the cases, the
parameters are all consistent within 1 $\sigma$ with their
$\Lambda$CDM values.

\begin{figure}
\centerline{\includegraphics[angle=-90,width=0.6\textwidth]{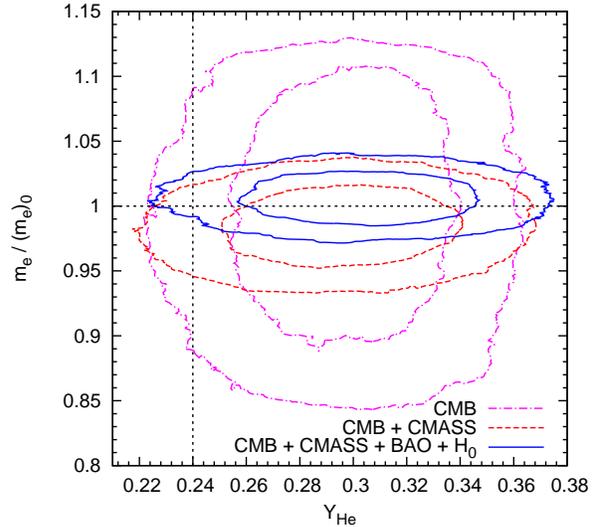}}
\caption{ The marginalized posterior distribution in the $m_e$ -
  $Y_{\rm He}$ plane for the $\Lambda$CDM parameter set extended to
  include the variation of $m_e$ and a variable primordial helium
  fraction $Y_{\rm He}$.  The dot-dashed lines show the 68\% and 95\%
  contours obtained using CMB information alone. The dashed lines
  correspond to the results obtained from the combination of CMB data
  plus the shape of the CMASS $\xi(s)$. The solid lines indicate the
  results obtained from the full dataset combination
  (CMB+CMASS+BAO+$H_0$). }
\label{fig:2D_me_YHe_variation_me-YHe}
\end{figure}

We now analyze the results for the case in which $m_e$ and $Y_{\rm
  He}$ are allowed to vary. Fig.~\ref{fig:2D_me_YHe_variation_me-YHe}
shows the two-dimensional marginalized constraints in the $Y_{\rm He}$
- $m_e$ plane. There is no degeneracy between $m_e$ and $Y_{\rm He}$
for any of the datasets.

\begin{table} 
\centering
  \caption{ The marginalized 68\% allowed regions on the cosmological
    parameters of the $\Lambda$CDM model, adding the variation of the
    electron mass $m_e$ and $Y_{\rm He}$ as a free parameter, obtained
    using different combinations of the datasets.  }
\begin{tabular}{@{}lccc@{}}
\hline
& \multirow{2}{*}{CMB}  & \multirow{2}{*}{CMB + CMASS} &  CMB + CMASS \\
&                       &                           &   + BAO + $H_0$   \\  
\hline
$Y_{\rm He}$  &  $0.296_{-0.029}^{+0.029}$ &  $0.295_{-0.030}^{+0.029}$ &  $0.302_{-0.029}^{+0.029}$ \\[1.5mm]
$m_e/(m_e)_0$  &  $0.994_{-0.068}^{+0.070}$ &  $0.985_{-0.021}^{+0.020}$ &  $1.006_{-0.014}^{+0.013}$ \\[1.5mm]
100$\Theta$  &  $1.037_{-0.050}^{+0.051}$ &  $1.031_{-0.015}^{+0.015}$ &  $1.0473_{-0.0095}^{+0.0094}$ \\[1.5mm]
100$\omega_{\rm b}$  &  $2.24_{-0.17}^{+0.17}$ &  $2.216_{-0.059}^{+0.058}$ &  $2.267_{-0.046}^{+0.046}$ \\[1.5mm]
100$\omega_{\rm dm}$  &  $11.33_{-0.92}^{+0.91}$ &  $11.20_{-0.66}^{+0.65}$ &  $11.87_{-0.56}^{+0.55}$ \\[1.5mm]
$\tau$  &  $0.0881_{-0.0075}^{+0.0064}$ &  $0.0882_{-0.0075}^{+0.0065}$ &  $0.0849_{-0.0070}^{+0.0063}$ \\[1.5mm]
$n_s$  &  $0.978_{-0.014}^{+0.014}$ &  $0.978_{-0.013}^{+0.013}$ &  $0.977_{-0.013}^{+0.013}$ \\[1.5mm]
ln$(10^{10}A_{\rm s})$  &  $3.106_{-0.035}^{+0.035}$ &  $3.105_{-0.034}^{+0.034}$ &  $3.116_{-0.033}^{+0.033}$ \\[1.5mm]
$\Omega_{\rm DE}$  &  $0.69_{-0.11}^{+0.11}$ &  $0.706_{-0.020}^{+0.020}$ &  $0.718_{-0.010}^{+0.010}$ \\[1.5mm]
$\Omega_{\rm m}$  &  $0.31_{-0.11}^{+0.11}$ &  $0.294_{-0.020}^{+0.020}$ &  $0.282_{-0.010}^{+0.010}$ \\[1.5mm]
$\sigma_{\rm 8}$  &  $0.827_{-0.079}^{+0.077}$ &  $0.821_{-0.042}^{+0.041}$ &  $0.863_{-0.035}^{+0.035}$ \\[1.5mm]
$t_{\rm 0}$/Gyr   &  $13.9_{-1.6}^{+1.6}$ &  $14.04_{-0.47}^{+0.47}$ &  $13.54_{-0.28}^{+0.28}$ \\[1.5mm]
$z_{\rm re}$  &  $10.9_{-1.5}^{+1.5}$ &  $10.8_{-1.2}^{+1.3}$ &  $11.9_{-1.3}^{+1.3}$ \\[1.5mm]
$h$  &  $0.70_{-0.15}^{+0.16}$ &  $0.677_{-0.033}^{+0.033}$ &  $0.709_{-0.017}^{+0.017}$ \\[1.5mm]
\hline
\end{tabular}
\label{tab:me_YHe}
\end{table}


In Table~\ref{tab:me_YHe} we present the constraints on $Y_{\rm He}$,
$m_e$, and the rest of the cosmological parameters for all of our
datasets.  For the CMB dataset, $Y_{\rm He} = 0.296 \pm 0.029$, which
is almost 2 $\sigma$ higher than the standard value.  The constraint
on $m_e$ is $m_e/(m_e)_0 = 0.994_{-0.068}^{+0.070}$. By adding the
CMASS correlation function, we tighten the bounds on $m_e$ by a factor
of three ($m_e /(m_e)_0 = 0.985_{-0.021}^{+0.020}$) but the bound on
$Y_{\rm He}$ is unchanged.  When all datasets are considered in the
analysis, the constraints are $Y_{\rm He}= 0.302 \pm 0.029$ and $m_e
/(m_e)_0 = 1.006_{-0.014}^{+0.013}$. The $H_0$ prior shifts $Y_{\rm
  He}$ towards slightly higher values, which are more than 2 $\sigma$
above the standard value of 0.24. On the other hand, $m_e$ is
consistent with no variation at 1 $\sigma$, for all of the datasets.

 Regarding the values of the remaining cosmological parameters, we
 find that there is a slight tension at the 1 $\sigma$ level for the
 parameter $n_{\rm s}$, which is higher than its $\Lambda$CDM value
 for the CMB+CMASS dataset, and for the parameter $\sigma_{\rm 8}$,
 which is higher than its $\Lambda$CDM value for the full dataset. In
 the rest of the cases, the parameters are all consistent within 1
 $\sigma$ with their $\Lambda$CDM values.

\section{Conclusions}
\label{sec:conclusions}

In this paper we have presented new constraints on the variation of
the fine structure constant and on the electron mass using the latest
CMB observations, and the full shape of the (spherically averaged)
redshift-space correlation function of the CMASS sample of galaxies,
drawn from the Data Release 9 (DR9) of the Baryonic Oscillations
Spectroscopic Survey (BOSS). Recent BAO and $H_0$ measurements were
also considered.  We have studied the degeneracies between these
constants and other cosmological parameters, such as the dark energy
equation of state, the neutrino mass, the effective number of
relativistic species, and the primordial helium abundance.  The main
results can be summarized as follows:

\begin{enumerate}

\item In the case of variation of only $\alpha$, our bound is $\alpha
  / \alpha_0 = 0.9957_{-0.0042}^{+0.0041}$, consistent with no
  variation at the 2 $\sigma$ level (almost at 1 $\sigma$). The
  constraints from CMB data alone are slightly improved when CMASS and
  other cosmological datasets are included.

\item When only $m_e$ is allowed to vary, the bounds are highly
  improved when additional datasets are considered in the
  analysis. Our best estimate including all datasets is $m_e / (m_e)_0
  = 1.006_{-0.013}^{+0.014}$, consistent with no variation of $m_e$
  within 1 $\sigma$. The CMASS dataset improves largely the CMB-only
  constraints.

\item When both $\alpha$ and $m_e$ are allowed to vary, the
  constraints on $\alpha$ do not improve with the addition of new
  datasets, while they do for $m_e$. With each dataset addition, the
  variation of the constants becomes more correlated. Our final
  bounds, using the complete dataset, are $\alpha / \alpha_0 =
  0.9901_{-0.0054}^{+0.0055}$ and $m_e / (m_e)_0 = 1.028 \pm
  0.019$. Both limits are consistent with no variation of the
  fundamental constants within 2 $\sigma$.

\item When we study the variation of $\alpha$ taking the value of
  $w_{\rm DE}$ as a free parameter, the allowed region for $\alpha$
  hardly depends on the addition of further datasets. The constraints
  on $w_{\rm DE}$, however, are noticeably improved. The bounds for
  the complete datasets are $\alpha/ \alpha_0 = 0.9915 \pm 0.0048$ and
  $w_{\rm DE} = -1.20 \pm 0.13$, which deviates from the standard
  value $w_{\rm DE} = -1$ at 1 $\sigma$.

\item When both $m_e$ and $w_{\rm DE}$ are allowed to vary, the CMB
  dataset cannot place tight constraints on both quantities.  The
  inclusion of CMASS correlation function helps to improve the
  bounds. Our results for the full dataset are $m_e / (m_e)_0 = 0.996
  \pm 0.029$ and $w_{\rm DE} = -1.12 \pm 0.23$.

\item When we allow for a non-negligible contribution of massive
  neutrinos to the dark matter component, our result is $\alpha/
  \alpha_0 = 0.9978_{-0.0045}^{+0.0044}$, and $f_\nu < 0.043$ at 95\%
  CL for the full dataset. The constraint on the sum of the neutrino
  masses is $\sum m_\nu < 0.46 \, {\rm eV}$ at 95\% CL.

 \item In the case of joint variation of $m_e$ and $f_\nu$, the bounds
   for the full dataset are $m_e / (m_e)_0 = 1.035 \pm 0.021$, and
   $f_\nu < 0.086$ at 95\% CL. The degeneracies between these
   quantities and $H_0$ shift the constraints to higher values when
   the HST $H_0$ prior is included.

 \item When $N_{\rm eff}$ and $\alpha$ are allowed to vary, the values
   of $N_{\rm eff}$ are consistent within 1 $\sigma$ with its standard
   value for the CMB and CMB+CMASS datasets. When the HST $H_0$ prior
   is added, the value of $N_{\rm eff}$ is much higher, being slightly
   consistent at 2 $\sigma$ with its standard value. The constraint on
   $\alpha$ is $\alpha/\alpha_0 = 0.9967 \pm 0.0042$, fully consistent
   with no variation at 1 $\sigma$.

 \item When $N_{\rm eff}$ and $m_e$ are allowed to vary, the value of
   $N_{\rm eff}$ is more than 1 $\sigma$ higher than its standard
   value, and the value of $m_e$ is consistent with no variation at 1
   $\sigma$, for all of the datasets.

\item In the case of joint variation of $Y_{\rm He}$ and $\alpha$,
  there is a strong degeneracy between these quantities, for all of
  the datasets. For the CMB and CMB+CMASS datasets, $Y_{\rm He}$ is
  consistent with the standard value of 0.24. When the $H_0$ prior is
  added, $Y_{\rm He}$ is almost 2 $\sigma$ higher than the standard
  value. For all of the datasets, there is no variation of $\alpha$ at
  the 1 $\sigma$ level.

\item In the case of joint variation of $Y_{\rm He}$ and $m_e$, there
  is no correlation between these quantities for any dataset. The
  bounds on $Y_{\rm He}$ are similar for all of the datasets, a value
  around 2 $\sigma$ higher than the standard one. For all of the
  datasets, $m_e$ is consistent with no variation at 1 $\sigma$.

\item We present new bounds on $N_{\rm eff}$ and $Y_{\rm He}$ in the
  case of no variation of fundamental constants. For the full dataset,
  in the $\Lambda$CDM+$N_{\rm eff}$ model, the constraint is $N_{\rm
    eff} = 3.86_{-0.40}^{+0.39}$. In the $\Lambda$CDM+$Y_{\rm He}$
  model, the constraint is $Y_{\rm He} = 0.301_{-0.029}^{+0.028}$.

\end{enumerate}

The analysis carried out in this paper is based on the first
spectroscopic data release of BOSS. Future data releases will provide
even more accurate views on the LSS clustering pattern by probing a
larger volume of the universe. Together with Planck satellite data to
be released in early 2013, this will enable to put more stringent
constraints on the variation of fundamental constants.

\appendix

\section{Constraints on the $\Lambda$CDM model} 
\label{sec:table_LCDM}

In Table~\ref{tab:LCDM}, we present our constraints on the cosmological
parameters of the $\Lambda$CDM model, for better comparison with the
results obtained in this paper. These are consistent with the
constraints given in \citet{Sanchez2012}.

\begin{table} 
\centering
  \caption{ The marginalized 68\% allowed regions on the cosmological
    parameters of the $\Lambda$CDM model, obtained using different
    combinations of the datasets. }
    \begin{tabular}{@{}lccc@{}}
    \hline
& \multirow{2}{*}{CMB}  & \multirow{2}{*}{CMB + CMASS} &  CMB + CMASS \\
&                       &                           &   + BAO + $H_0$   \\  
\hline
100$\Theta$  &  $1.0411_{-0.0016}^{+0.0016}$ &  $1.0407_{-0.0015}^{+0.0015}$ &  $1.0410_{-0.0015}^{+0.0015}$ \\[1.5mm]
100$\omega_{\rm b}$  &  $2.223_{-0.042}^{+0.042}$ &  $2.212_{-0.038}^{+0.037}$ &  $2.223_{-0.038}^{+ 0.038}$ \\[1.5mm]
100$\omega_{\rm dm}$  &  $11.18_{-0.48}^{+ 0.48}$ &  $11.45_{-0.29}^{+0.29}$ &  $11.45_{-0.21}^{+ 0.21}$ \\[1.5mm]
$\tau$  &  $0.0850_{-0.0068}^{+0.0065}$ &  $0.0820_{-0.0066}^{+0.0057}$ &  $0.0826_{-0.0064}^{+0.0059}$ \\[1.5mm]
$n_{\rm s}$  &  $0.966_{-0.011}^{+0.011}$ &  $0.9620_{-0.0091}^{+0.0091}$ &  $0.9641_{-0.0087}^{+0.0088}$ \\[1.5mm]
ln$(10^{10}A_{\rm s})$  &  $3.081_{-0.030}^{+0.030}$ &  $3.084_{-0.028}^{+0.028}$ &  $3.086_{-0.028}^{+0.028}$ \\[1.5mm]
$\Omega_{\rm DE}$  &  $0.733_{-0.025}^{+0.025}$ &  $0.718_{-0.015}^{+0.015}$ &  $0.7193_{-0.0099}^{+0.0099}$ \\[1.5mm]
$\Omega_{\rm m}$  &  $0.267_{-0.025}^{+0.025}$ &  $0.282_{-0.015}^{+0.015}$ &  $0.2807_{-0.0099}^{+0.0099}$ \\[1.5mm]
$\sigma_{\rm 8}$  &  $0.814_{-0.024}^{+0.024}$ &  $0.825_{-0.018}^{+0.018}$ &  $0.827_{-0.016}^{+0.016}$ \\[1.5mm]
$t_{\rm 0}$/Gyr   &  $13.729_{-0.090}^{+0.089}$ &  $13.769_{-0.071}^{+0.071}$ &  $13.750_{-0.064}^{+0.065}$ \\[1.5mm]
$z_{\rm re}$  &  $10.4_{-1.2}^{+1.2}$ &  $10.2_{-1.2}^{+1.1}$ &  $10.3_{-1.1}^{+1.2}$ \\[1.5mm]
$h$  &  $0.710_{-0.021}^{+0.021}$ &  $0.697_{-0.012}^{+0.012}$ &  $0.6982_{-0.0083}^{+0.0085}$ \\[1.5mm]
 \hline
\end{tabular}
\label{tab:LCDM}
\end{table}


\section*{Acknowledgements}

CGS, JAR-M, RG-S, and RR acknowledge funding from project
AYA2010-21766-C03-02 of the Spanish Ministry of Science and Innovation
(MICINN). JAR-M is a Ram\'on y Cajal fellow of the Spanish Ministry of
Science and Innovation (MICINN).

Funding for SDSS-III has been provided by the Alfred P. Sloan
Foundation, the Participating Institutions, the National Science
Foundation, and the U.S. Department of Energy.  SDSS-III is managed by
the Astrophysical Research Consortium for the Participating
Institutions of the SDSS-III Collaboration including the University of
Arizona, the Brazilian Participation Group, Brookhaven National
Laboratory, University of Cambridge, University of Florida, the French
Participation Group, the German Participation Group, the Instituto de
Astrofisica de Canarias, the Michigan State/Notre Dame/JINA
Participation Group, Johns Hopkins University, Lawrence Berkeley
National Laboratory, Max Planck Institute for Astrophysics, Max Planck
Institute for Extraterrestrial Physics, New Mexico State University,
New York University, Ohio State University, Pennsylvania State
University, University of Portsmouth, Princeton University, the
Spanish Participation Group, University of Tokyo, University of Utah,
Vanderbilt University, University of Virginia, University of
Washington, and Yale University.

We acknowledge the use of the Legacy Archive for Microwave Background 
Data Analysis (LAMBDA). Support for LAMBDA is provided by the NASA
Office of Space Science.

\appendix

\end{document}